\pdfoutput=1
\documentclass[twocolumn, floatfix]{aastex63}

\usepackage[T1]{fontenc}
\usepackage{float}
\usepackage{xcolor}

\usepackage{ulem}

\makeatletter
\DeclareRobustCommand{\HI}{  \mbox{H\check@mathfonts\fontsize\sf@size\z@\selectfont I}}
\makeatother

\newcommand{\lya}{Ly$\alpha$}
\newcommand{\lyb}{Ly$\beta$}

\newcommand{\hi}{\ion{H}{1}}

\newcommand{\lcdm}{$\rm \Lambda$CDM\ }

\def\cmpch{~h^{-1}\,{\rm Mpc}} 
\def\teff{\tau_{\rm eff} }
\def\zem{z_{\rm em}}
\def\zabs{z_{\rm abs}}

\def\add#1{{\added{#1}}}

\submitjournal{\apj}

\shorttitle{Dark Gap Statistics}
\shortauthors{Zhu et al.}

\begin{document}

\title{Chasing the Tail of Cosmic Reionization with Dark Gap Statistics in the \lya\ Forest over $5 < z < 6$}

\correspondingauthor{Yongda Zhu}
\email{yzhu144@ucr.edu}

\author[0000-0003-3307-7525]{Yongda Zhu}
\affiliation{Department of Physics \& Astronomy,
    University of California, Riverside, CA 92521, USA}

\author[0000-0003-2344-263X]{George D. Becker}
\affiliation{Department of Physics \& Astronomy,
    University of California, Riverside, CA 92521, USA}

\author[0000-0001-8582-7012]{Sarah E. I. Bosman}
\affiliation{Max-Planck-Institut f\"{u}r Astronomie, K\"{o}nigstuhl 17, D-69117 Heidelberg, Germany}

\author[0000-0001-5211-1958]{Laura C. Keating}
\affiliation{Leibniz-Institut f\"ur Astrophysik Potsdam (AIP), An der Sternwarte 16, D-14482 Potsdam, Germany}

\author[0000-0002-0421-065X]{Holly M. Christenson}
\affiliation{Department of Physics \& Astronomy,
    University of California, Riverside, CA 92521, USA}

\author[0000-0002-2931-7824]{Eduardo Ba\~nados}
\affiliation{Max-Planck-Institut f\"{u}r Astronomie, K\"{o}nigstuhl 17, D-69117 Heidelberg, Germany}

\author[0000-0002-1620-0897]{Fuyan Bian}
\affiliation{European Southern Observatory, Alonso de Córdova 3107, Casilla 19001, Vitacura, Santiago 19, Chile}

\author[0000-0003-0821-3644]{Frederick B.~Davies}
\affiliation{Max-Planck-Institut f\"{u}r Astronomie, K\"{o}nigstuhl 17, D-69117 Heidelberg, Germany}

\author[0000-0003-3693-3091]{Valentina D'Odorico}
\affiliation{INAF-Osservatorio Astronomico di Trieste, Via Tiepolo 11, I-34143 Trieste, Italy}
\affiliation{Scuola Normale Superiore, Piazza dei Cavalieri 7, I-56126 Pisa, Italy}
\affiliation{IFPU—Institute for Fundamental Physics of the Universe, via Beirut 2, I-34151 Trieste, Italy}

\author[0000-0003-2895-6218]{Anna-Christina Eilers}\thanks{NASA Hubble Fellow}
\affiliation{MIT Kavli Institute for Astrophysics and Space Research, 77 Massachusetts Ave., Cambridge, MA 02139, USA}

\author[0000-0003-3310-0131]{Xiaohui Fan}
\affiliation{Steward Observatory, University of Arizona, 933 North Cherry Avenue, Tucson, AZ 85721, USA}

\author[0000-0001-8443-2393]{Martin G. Haehnelt}
\affiliation{Kavli Institute for Cosmology and Institute of Astronomy, Madingley Road, Cambridge, CB3 0HA, UK}

\author[0000-0001-5829-4716]{Girish Kulkarni}
\affiliation{Tata Institute of Fundamental Research, Homi Bhabha Road, Mumbai 400005, India}

\author[0000-0002-7129-5761]{Andrea Pallottini}
\affiliation{Scuola Normale Superiore, Piazza dei Cavalieri 7, I-56126 Pisa, Italy}

\author[0000-0002-4314-1810]{Yuxiang Qin}
\affiliation{School of Physics, University of Melbourne, Parkville, VIC 3010, Australia}
\affiliation{ARC Centre of Excellence for All Sky Astrophysics in 3 Dimensions (ASTRO 3D)}

\author[0000-0002-7633-431X]{Feige Wang}\thanks{NASA Hubble Fellow}
\affiliation{Steward Observatory, University of Arizona, 933 North Cherry Avenue, Tucson, AZ 85721, USA}

\author[0000-0001-5287-4242]{Jinyi Yang}\thanks{Strittmatter Fellow}
\affiliation{Steward Observatory, University of Arizona, 933 North Cherry Avenue, Tucson, AZ 85721, USA}

\begin{abstract}
    We present a new investigation of the intergalactic medium (IGM) near the end of reionization using ``dark gaps'' in the Lyman-alpha (\lya) forest.  Using spectra of 55 QSOs at $z_{\rm em}>5.5$, including new data from the XQR-30 VLT Large Programme, we identify gaps in the \lya\ forest where the transmission averaged over 1 comoving $\cmpch$ bins falls below 5\%. Nine ultra-long ($L > 80\cmpch$) dark gaps are identified at $z<6$. 
    In addition, we quantify the fraction of QSO spectra exhibiting gaps longer than $30\cmpch$,
        $F_{30}$, as a function of redshift.  We measure $F_{30} \simeq 0.9$, 0.6, and 0.15 at $z = 6.0$, 5.8, and 5.6, respectively, with the last of these long dark gaps persisting down to $z \simeq 5.3$.
    Comparing our results with predictions from hydrodynamical simulations,  we find that the data are consistent with  models wherein reionization extends significantly below redshift six. 
    Models wherein the IGM is essentially fully reionized 
        that retain large-scale fluctuations in the ionizing UV background at $z \lesssim 6$ are also potentially consistent with the data. Overall, our results suggest that signature of reionization in the form of islands of neutral hydrogen and/or large-scale fluctuations in the ionizing background remain present in the IGM until at least $z \simeq 5.3$.
\end{abstract}

\keywords{\href{http://astrothesaurus.org/uat/1383}{Reionization (1383)}, 
\href{http://astrothesaurus.org/uat/813}{Intergalactic medium (813)}, 
\href{http://astrothesaurus.org/uat/1317}{Quasar absorption line spectroscopy (1317)}, \href{http://astrothesaurus.org/uat/734}{High-redshift galaxies (734)}}

\section{Introduction}\label{sec: intro}
The reionization of the intergalactic medium (IGM) is 
the last major phase transition in the history of the Universe.
In the widely accepted picture, neutral hydrogen in the IGM 
was reionized by ultraviolet photons emitted by the first luminous sources \citep[e.g.,][]{bromm_first_2004,mcquinn_evolution_2016,dayal_early_2018}.
Determining when reionization occurred as well as what sources were responsible is therefore important for the understanding of formation and evolution of the first stars, galaxies, and black holes.

Multiple observations now constrain the timing of reionization.
Cosmic microwave background (CMB) measurements suggest a midpoint at redshift $z_{\rm re}=7.7\pm0.7$ (\citealp{planck_collaboration_planck_2020}; see also \citealp{de_belsunce_inference_2021}). The redshift evolution in the fraction of UV-selected galaxies detected in Lyman-alpha (\lya) emission also suggests that the IGM was significantly neutral near $z\sim7$--8 \citep[e.g.,][and references therein, but see \citealt{wold_lager_2021-1}]{mason_universe_2018,mason_inferences_2019,hoag_constraining_2019,hu_ly_2019}.
These results are broadly consistent with multiple probes of the IGM using QSO spectra.
For example, the IGM thermal history at $z > 4$ inferred from the \lya\ flux power spectrum suggests a mean redshift of reionization near $z_{\rm re}\simeq8.5_{-0.8}^{+1.1}$ (\citealp{boera_revealing_2019}, see also \citealp{gaikwad_consistent_2020,walther_new_2019}).
Similarly, \lya\ damping wing measurements of $\zem >7$ QSOs indicate that the IGM was significantly neutral at $z\sim7$--7.5 \citep[e.g.,][]{banados_800-million-solar-mass_2018,davies_quantitative_2018,greig_are_2017,greig_constraints_2019,wang_significantly_2020-1,yang_poniuaena_2020}.
The appearance of transmitted flux in the \lya\ and \lyb\ forests suggests that the reionization mostly completed by $z\simeq6$ \citep[e.g.,][]{mcgreer_model-independent_2015}. On the other hand, large fluctuations in the observed IGM effective opacity ($\teff = -\ln{\langle F \rangle}$, where $F$ is the continuum-normalized flux) in the \lya\ forest
at $z < 6$ suggest that signatures of reionization may persist in the IGM down to even lower redshifts \citep[][]{fan_constraining_2006,becker_evidence_2015,
bosman_new_2018,eilers_opacity_2018,yang_measurements_2020-1,bosman_hydrogen_2021}.

Multiple models have been proposed to explain the large-scale fluctuations in 
IGM \lya\ opacity at $z < 6$.  If the IGM is mostly ionized at $z \ge 6$, then large variations in \lya\ opacity may persist to lower redshifts due to either lingering temperature fluctuations produced by inhomogeneous reionization \citep{daloisio_large_2015}, or fluctuations in the ionizing UV background produced by a short and spatially variable mean free path (\citealp{davies_large_2016,nasir_observing_2020}), or rare sources such as QSOs (\citealp{chardin_large-scale_2017}; see also \citealp{meiksin_influence_2020}).
Alternatively, if reionization continues substantially to $z < 6$ then the observed scatter in $\teff$ could be due to the presence of large patches of neutral gas coupled with UVB fluctuations \citep[e.g.,][]{kulkarni_large_2019,keating_constraining_2020,nasir_observing_2020,qin_reionization_2021}.
The combination of neutral patches and UVB fluctuations may naturally explain the presence of giant \lya\ troughs such as the $110\cmpch$ trough towards ULAS J0148+0600 identified by \citet{becker_evidence_2015} \citep[e.g.,][]{keating_long_2020}. 
A late-ending reionization\footnote{
Throughout this paper, for convenience, we refer to ``the end of reionization'' as when 
the volume filling factor of ionized gas in the IGM reaches 99\%. We use ``early'' for scenarios wherein reionization ends at $z\geq6$, 
and ``late'' for cases where reionization ends below $z=6$.}
scenario is also consistent with the evolution of \ion{O}{1} absorbers at $z\simeq6$ \citep[][]{becker_evolution_2019},  and is preferred by recent Bayesian inference results that simultaneously match \lya\ forest, CMB, and galaxy data \citep{choudhury_studying_2021, qin_reionization_2021}.

Some progress has been made towards distinguishing between these models observationally. Measurements of \lya\ emitting galaxies (LAEs, 
\citealp{becker_evidence_2018}, 
Christenson et al., in prep) and Lyman break galaxies (LBGs, 
\citealp{kashino_evidence_2020}) in the field of ULAS J0148+0600 have demonstrated that the \lya\ trough along this line of sight is associated with a large-scale underdensity.  This result disfavors the temperature fluctuation model, but is consistent with either the pure UVB fluctuation or late-reionization model.  The QSO UVB model is potentially also consistent with this result, though it is disfavored by measurements of the QSO luminosity function near $z \sim 6$ \citep[e.g.,][]{parsa_no_2018,kulkarni_evolution_2019}. Recent measurements of the mean free path of ionizing photons over $5<z<6$ are consistent with late reionization models wherein the IGM is still significantly neutral at $z=6$, and disfavor models in which reionization ends early enough that the IGM relaxes hydrodynamically by $z=6$ \citep{becker_mean_2021}.  Simultaneously matching the mean free path measurements and other IGM constraints further favors a late and rapid reionization scenario driven by galaxies that are efficient at producing and/or emitting ionizing photons \citep{cain_short_2021,davies_predicament_2021}.

A key question, therefore, is whether and for how long the impacts of reionization continued below $z = 6$.  It is also of interest to determine whether a late reionization scenario with islands of neutral gas and UVB fluctuations at $z < 6$ can be distinguished observationally from an early reionization scenario with UVB fluctuations alone. 
Better measurements of the spatial scale of the \lya\ opacity fluctuations may provide some insights.  Determining how long these fluctuations persist and how they evolve with redshift may also be helpful.
One way to do this is by identifying individual ``dark gaps'' in the \lya\ forest \citep[e.g.][]{songaila_approaching_2002,furlanetto_constraining_2004,paschos_statistical_2005,fan_constraining_2006,gallerani_glimpsing_2008}, which could be created by regions of neutral IGM and/or low UV background \citep[e.g.,][]{nasir_observing_2020}. 
Because dark gaps contain multi-scale spatial information,
they provide complementary information to $\teff$ measurements averaged over 
intervals of fixed length, 
and may therefore be useful for distinguishing between models of the IGM at $z < 6$.

In this paper, we use dark gap statistics to characterize the opacity of the IGM over $5\lesssim z \lesssim 6$. 
In particular, we use these statistics to determine how long large \lya-opaque regions persist in the IGM, and whether the data are
consistent with existing late reionization models and/or early reionization models that retain a fluctuating UVB.  We use a sample of 55 high signal-to-noise (S/N) spectra of QSOs at 
$5.5\lesssim \zem \lesssim 6.5$, including 23 new X-Shooter spectra from the XQR-30 VLT Large Programme (D'Odorico et al., in prep).
In addition to the distribution of dark gap lengths, we 
measure the fraction of QSO spectra exhibiting long ($L \ge 30\cmpch$) dark gaps as a function of
redshift for the first time.

We present our data in Section \ref{sec:data}.
In Section \ref{sec:dark_gap_stat} we describe our methods for measuring dark gaps
and the results of dark gap statistics.
Section \ref{sec:simulations} introduces the 
models to which we compare our measurements.
We then discuss the implications for the models in 
Section \ref{sec:discussion}.
Finally, we summarize our conclusions in Section \ref{SEC:SUMMARY}.
Throughout this paper we quote distances in comoving units unless otherwise noted, and 
assume a \lcdm cosmology with $h = 0.678$, $\Omega_m=0.308$ and
$\Omega_{\rm \Lambda}=0.692$.

\section{The data \label{sec:data}}
\subsection{QSO spectra}

This study is based on spectra of 55 QSOs at $5.5\lesssim \zem \lesssim 6.5$ 
taken with the X-Shooter spectrograph on the Very Large Telescope \citep[VLT;][]{vernet_x-shooter_2011} and the Echellette Spectrograph and Imager (ESI) on Keck \citep{sheinis_esi_2002}.
Of these, 23 X-Shooter spectra are from the XQR-30 VLT Large Programme.
\add{The XQR-30 program is targeting 30} bright QSOs at $5.8\lesssim z \lesssim 6.6$ for the study of reionization and other aspects of the early Universe. The full data set will be described in D'Odorico et al., in prep. \add{The 23 objects out of the XQR-30 sample selected for this project are those that meet our S/N threshold and do not contain strong BAL features.}  \add{In addition, we} use 30 spectra reduced from archival X-Shooter and ESI data, of which 27 are from the sample of \citet{becker_evolution_2019}. Recent deep (20 hour) X-Shooter observations (PI: Fuyan Bian) of the lensed $z=6.5$ QSO \add{J0439+1634} are also included in the dark gaps statistics. 
Finally, we acquired a deep (7 hour) ESI spectrum of SDSS J1250+3130.  Observations for all objects except SDSS J1250+3130 were taken without any foreknowledge of dark gaps in the \lya\ forest.  In the case of SDSS J1250+3130, we targeted the QSO based on indications from a shallower (1 hour) ESI spectrum that its spectrum contained a long dark gap in the \lya\ forest.  We discuss the impact of including this object on our results in Section \ref{sec:F30}.

\begin{figure*}
\centering
\includegraphics[width=6.5in]{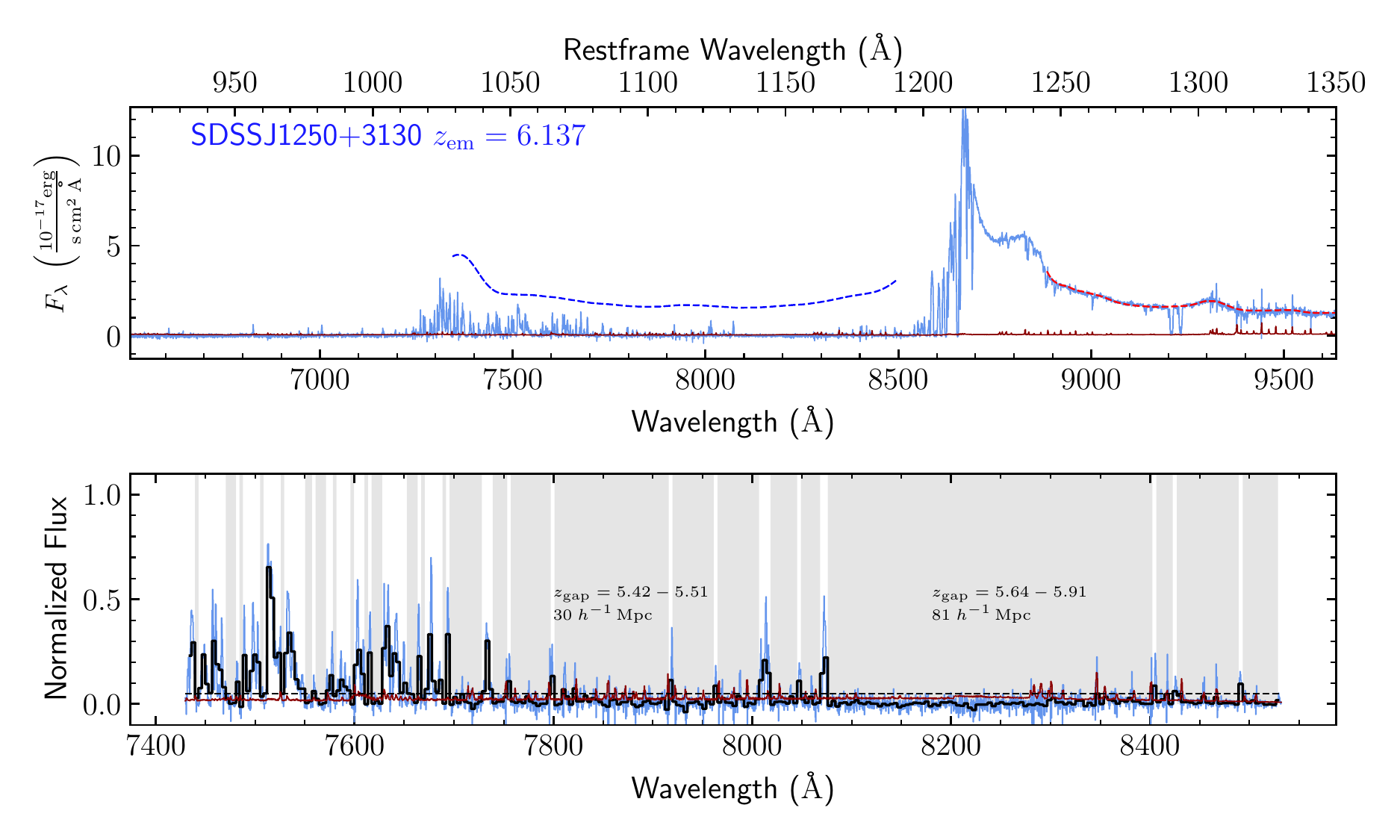}
\caption{Spectrum, continuum fits and dark gap detection details
of the $\zem=6.137$ quasar SDSS J1250+3130. 
{\bf Top}: QSO spectrum and continuum fits. The light blue and dark red lines represents flux and flux error in the original binning. Dashed red and blue curves are the best-fitting QSO continuum based on PCA. {\bf Bottom}: \lya\ forest and dark gaps detected. The dashed black line labels the flux threshold of 0.05. The thick black line displays the flux binned to $1\cmpch$. Light blue and dark red lines show the flux and flux error in the original binning. Dark gaps detected are shaded with gray. We also label the redshift range and length of each long dark gap ($L\geq30\cmpch$), if any. \\(The complete figure set (55 images) is available in the online journal. For this preprint, online materials are available at \url{https://ydzhuastro.github.io/Zhu21.html}.)}
\label{fig:appendix01}
\end{figure*}

Details of the data reduction are given in \citet{becker_evolution_2019}.  Briefly, we used a custom pipeline that includes optimal techniques for sky subtraction \citep{kelson_optimal_2003} and one-dimensional spectral extraction \citep{horne_optimal_1986}. Telluric absorption corrections were computed for individual exposures using models based on the Cerro Paranal Advanced Sky Model  \citep{noll_atmospheric_2012,jones_advanced_2013}. The spectra were extracted using 10 ${\rm km~s^{-1}}$ pixels for the VIS arm of X-Shooter and 15 ${\rm km~s^{-1}}$ pixels for ESI.  Typical resolutions for the X-Shooter and ESI are ${\rm FWHM} \approx 25~{\rm km~s^{-1}}$ and 45 ${\rm km~s^{-1}}$, respectively. In addition, for \add{J0439+1634}, to reduce the contamination from continuum emission of a foreground galaxy, we fit a power law of the flux zero point over the \lya\ forest and subtract it from the flux. The spectra are plotted in Figure Set~\ref{fig:appendix01}.

\startlongtable
\begin{deluxetable*}{cllccr}
    \tablenum{1}
    \tablecaption{QSO spectra used in this work \label{tab:QSOlist}}
    \tablehead{
        \colhead{No.} & \colhead{QSO} & \colhead{$z_{\rm em}^{\rm [Ref.]}$} &
        \colhead{Source} & \colhead{Instrument} & \colhead{S/N}
    }
    \decimalcolnumbers
    \startdata
1  & J2207-0416             & $5.529  ^{\rm b}$ & archival (B19)   & X-Shooter & 42  \\
2  & J0108+0711             & $5.577  ^{\rm b}$ & archival (B19)   & X-Shooter & 29  \\
3  & J1335-0328             & $5.693  ^{\rm b}$ & archival (B19)   & X-Shooter & 30  \\
4  & SDSSJ0927+2001         & $5.7722 ^{\rm c}$ & archival (B19)   & X-Shooter & \add{76}  \\
5  & SDSSJ1044-0125         & $5.7847 ^{\rm o}$ & other archival   & ESI       & 71  \\
6  & PSOJ065+01             & $5.790  ^{\rm q}$ & XQR-30           & X-Shooter & \add{47}  \\
7  & PSOJ308-27             & $5.794  ^{\rm q}$ & XQR-30           & X-Shooter & \add{58}  \\
8  & SDSSJ0836+0054         & $5.810  ^{\rm g}$ & other archival   & ESI       & 152  \\
9  & PSOJ004+17             & $5.8165 ^{\rm e}$ & other archival   & X-Shooter & 21  \\
10 & SDSSJ0002+2550         & $5.820  ^{\rm b}$ & archival (B19)   & ESI       & 93  \\
11 & PSOJ242-12             & $5.834  ^{\rm q}$ & XQR-30           & X-Shooter & \add{28}  \\
12 & SDSSJ0840+5624         & $5.8441 ^{\rm n}$ & archival (B19)   & ESI       & 41  \\
13 & SDSSJ0005-0006         & $5.847  ^{\rm b}$ & archival (B19)   & ESI       & 24  \\
14 & PSOJ025-11             & $5.849  ^{\rm q}$ & XQR-30           & X-Shooter & \add{53}  \\
15 & PSOJ183-12             & $5.857  ^{\rm q}$ & XQR-30           & X-Shooter & \add{66}  \\
16 & SDSSJ1411+1217         & $5.904  ^{\rm g}$ & archival (B19)   & ESI       & 46  \\
17 & PSOJ108+08             & $5.950  ^{\rm q}$ & XQR-30           & X-Shooter & \add{70}  \\
18 & PSOJ056-16             & $5.9670 ^{\rm e}$ & archival (B19)   & X-Shooter & 35  \\
19 & PSOJ029-29             & $5.981  ^{\rm q}$ & XQR-30           & X-Shooter & \add{51}  \\
20 & SDSSJ0818+1722         & $5.997  ^{\rm b}$ & archival (B19)   & X-Shooter & \add{108}  \\
21 & ULASJ0148+0600         & $5.998  ^{\rm b}$ & archival (B19)   & X-Shooter & \add{126}  \\
22 & PSOJ340-18             & $5.999  ^{\rm b}$ & archival (B19)   & X-Shooter & 32  \\
23 & PSOJ007+04             & $6.0008 ^{\rm d}$ & XQR-30           & X-Shooter & \add{53}  \\
24 & SDSSJ2310+1855         & $6.0031 ^{\rm o}$ & XQR-30           & X-Shooter & \add{81}  \\
25 & SDSSJ1137+3549         & $6.007  ^{\rm j}$ & archival (B19)   & ESI       & 28  \\
26 & ATLASJ029.9915-36.5658 & $6.021  ^{\rm b}$ & XQR-30           & X-Shooter & \add{48}  \\
27 & SDSSJ1306+0356         & $6.0330 ^{\rm k}$ & archival (B19)   & X-Shooter & \add{71}  \\
28 & J0408-5632             & $6.035  ^{\rm q}$ & XQR-30           & X-Shooter & \add{71}  \\
29 & ULASJ1207+0630         & $6.0366 ^{\rm d}$ & archival (B19)   & X-Shooter & 25  \\
30 & SDSSJ2054-0005         & $6.0391 ^{\rm o}$ & archival (B19)   & ESI       & 29  \\
31 & PSOJ158-14             & $6.0681 ^{\rm e}$ & XQR-30           & X-Shooter & \add{59}  \\
32 & SDSSJ0842+1218         & $6.0763 ^{\rm d}$ & XQR-30           & X-Shooter & \add{71}  \\
33 & SDSSJ1602+4228         & $6.079  ^{\rm j}$ & archival (B19)   & ESI       & 34  \\
34 & PSOJ239-07             & $6.1098 ^{\rm e}$ & XQR-30           & X-Shooter & \add{65}  \\
35 & CFHQSJ1509-1749        & $6.1225 ^{\rm d}$ & archival (B19)   & X-Shooter & \add{54}  \\
36 & SDSSJ2315-0023         & $6.124  ^{\rm b}$ & archival (B19)   & ESI       & 25  \\
37 & ULASJ1319+0950         & $6.1330 ^{\rm o}$ & archival (B19)   & X-Shooter & \add{86}  \\
38 & SDSSJ1250+3130         & $6.137  ^{\rm j}$ & new observation  & ESI       & 53  \\
39 & VIKJ2318-3029          & $6.1458 ^{\rm d}$ & archival (B19)   & X-Shooter & 21  \\
40 & PSOJ217-16             & $6.1498 ^{\rm d}$ & XQR-30           & X-Shooter & \add{68}  \\
41 & PSOJ217-07             & $6.165  ^{\rm q}$ & XQR-30           & X-Shooter & \add{42}  \\
42 & PSOJ359-06             & $6.1718 ^{\rm e}$ & XQR-30           & X-Shooter & \add{67}  \\
43 & PSOJ060+24             & $6.177  ^{\rm q}$ & XQR-30           & X-Shooter & \add{53}  \\
44 & PSOJ065-26             & $6.1877 ^{\rm d}$ & XQR-30           & X-Shooter & \add{73}  \\
45 & PSOJ308-21             & $6.2341 ^{\rm d}$ & archival (B19)   & X-Shooter & 26  \\
46 & SDSSJ1030+0524         & $6.309  ^{\rm f}$ & archival (B19)   & X-Shooter & \add{35}  \\
47 & SDSSJ0100+2802         & $6.3270 ^{\rm l}$ & archival (B19)   & X-Shooter & \add{212}  \\
48 & ATLASJ025.6821-33.4627 & $6.3373 ^{\rm k}$ & archival (B19)   & X-Shooter & \add{61}  \\
49 & J1535+1943             & $6.381  ^{\rm q}$ & XQR-30           & X-Shooter & \add{30}  \\
50 & SDSSJ1148+5251         & $6.4189 ^{\rm h}$ & archival (B19)   & ESI       & 64  \\
51 & J1212+0505             & $6.4386 ^{\rm d}$ & XQR-30           & X-Shooter & \add{41}  \\
52 & J0439+1634             & $6.5188 ^{\rm p}$ & new observation  & X-Shooter & \add{224}  \\
53 & VDESJ0224-4711         & $6.5223 ^{\rm m}$ & XQR-30           & X-Shooter & \add{29}  \\
54 & PSOJ036+03             & $6.541  ^{\rm a}$ & archival (B19)   & X-Shooter & \add{38}  \\
55 & PSOJ323+12             & $6.5881 ^{\rm i}$ & XQR-30           & X-Shooter & \add{30}  \\
    \enddata
    \tablecomments{Columns: (1) QSO index number,
    (2) QSO name, (3) QSO redshift with reference, 
    (4) source of the spectrum used for dark gap statistics,
    \add{(5)} instrument used for dark gap statistics,
    \add{(6)} continuum signal-to-noise ratio per $30~{\rm km~s^{-1}}$ near rest wavelength 1285 \AA.\\
    Sources of the spectra. XQR-30: spectra from the XQR-30 program; new observation: spectra from new observations; archival (B19): archival spectra used and reduced in \citet{becker_evolution_2019}; other archival: spectra from the public archives but not included in \citet{becker_evolution_2019}.
    }

    \tablerefs{ Redshift lines and references. 
    a. [\ion{C}{2}] 158$\mu$m: \citet{banados_bright_2015};
    b. apparent start of the \lya\ forest: \citet{becker_evolution_2019};
    c. CO: \citet{carilli_detection_2007};
    d. [\ion{C}{2}] 158$\mu$m: \citet{decarli_alma_2018};
    e. [\ion{C}{2}] 158$\mu$m: \citet{eilers_detecting_2020};
    f. \ion{Mg}{2}: \citet{jiang_gemini_2007};
    g. \ion{Mg}{2}: \citet{kurk_black_2007};
    h. [\ion{C}{2}] 158$\mu$m: \citet{maiolino_first_2005};
    i. [\ion{C}{2}] 158$\mu$m: \citet{mazzucchelli_physical_2017};
    j. \ion{Mg}{2}: \citet{shen_gemini_2019};
    k. [\ion{C}{2}] 158$\mu$m: \citet{venemans_kiloparsec-scale_2020-1};
    l. [\ion{C}{2}] 158$\mu$m: \citet{wang_spatially_2019};
    m. [\ion{C}{2}] 158$\mu$m: \citet{wang_revealing_2021};
    n. CO: \citet{wang_molecular_2010};
    o. [\ion{C}{2}] 158$\mu$m: \citet{wang_star_2013};
    p. [\ion{C}{2}] 158$\mu$m: \citet{yang_far-infrared_2019};
    q. apparent start of the \lya\ forest: this work.
                                }
\end{deluxetable*}

\figsetstart
\figsetnum{1}
\figsettitle{QSO spectra and continuum fits}
\figsetgrpstart
\figsetgrpnum{1.38}
\figsetgrptitle{SDSSJ1250+3130}
\figsetplot{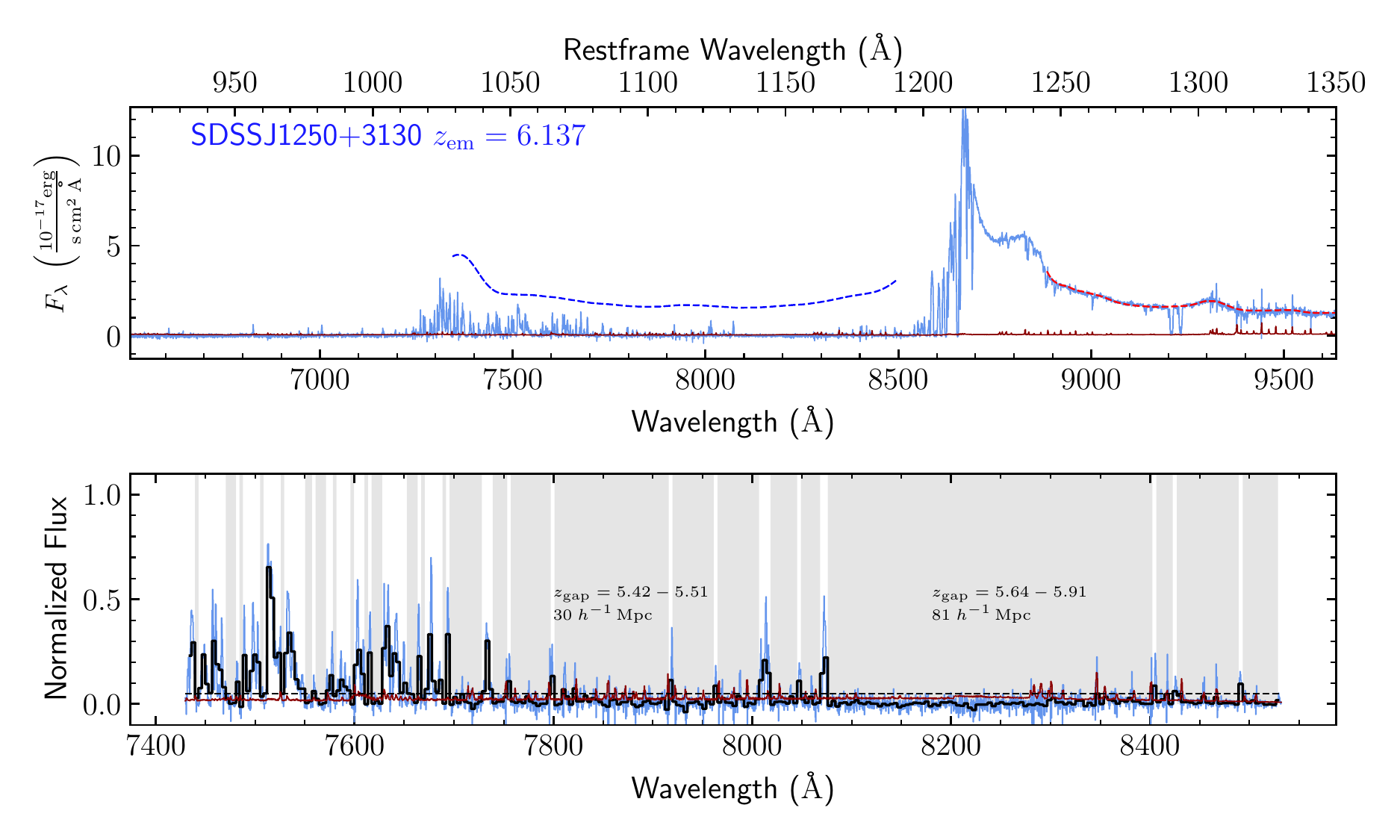}
\figsetgrpnote{Spectrum, continuum fitting and dark gap detection details
        of $\zem=6.137$ quasar SDSSJ1250+3130. 
        Lines are as described in Figure \ref{fig:appendix01}.}
\figsetgrpend
\figsetend

We adopt QSO redshifts measured from CO, [\ion{C}{2}] 158$\mu$m or \ion{Mg}{2} lines if available. Otherwise we use redshifts inferred from the apparent start of the \lya\ forest, following \citet{becker_evolution_2019}. Table \ref{tab:QSOlist} summarizes QSO spectra used in this work with the QSO redshifts, instruments, and estimated signal-to-noise ratios, which is calculated as the median ratio of unabsorbed QSO continuum to noise per $30~{\rm km~s^{-1}}$ near 1285 \AA\ in the rest frame.

\subsection{Continuum Fitting \label{sec:fitting}}
The detection of dark gaps relies on the construction of the intrinsic 
continuum over the \lya\ forest. In order to estimate QSO continua blueward of the \lya\ emission line, we use Principal Component Analysis (PCA), which is less biased 
than the conventional power law fitting \citep[e.g.,][]{bosman_comparison_2021}.
In this work, we apply the log-PCA method of \cite{davies_predicting_2018} as implemented in the \lya\ forest portion of the spectrum by \citet{bosman_hydrogen_2021},
with 15 and 10 components used for the red-side (rest-frame wavelength $\lambda_0>1230$ \AA) and blue-side ($\lambda_0<1170$ \AA) continuum. 
For each QSO, we fit the red-side continuum with principal components, and map the corresponding red principal component coefficients to the blue side coefficients with a projection matrix.
For X-Shooter spectra with observations from the NIR arm, we fit the red continuum over $1230<\lambda<2000$ \AA\ in the rest frame. 

The ESI spectra are fit using an optical-only PCA, which is presented in \citet{bosman_comparison_2021}. QSOs with strong broad absorption lines (BALs) in their spectra were excluded from our sample.
For QSOs with mild absorption features that interfere minimally with the \lya\ forest, we mask out the absorption lines when fitting their spectra. In addition, we intentionally leave out the \lya\ emission peak and the proximity zone when fitting and predicting the continuum on account of the large object-to-object variations in these regions. 
\add{The typical $1\sigma$ uncertainty of the PCA continuum fitting over the \lya\ forest is less than 10\%.}
Continuum fits and blue-side predictions are shown in Figure Set \ref{fig:appendix01} along with the QSO spectra. \add{We also verify that our dark gap statistics results do not significant change if we use power-law continua (see Appendix \ref{app:powerlaw}), which have a typical bias of $\sim$10\% over the \lya\ forest \citep{bosman_comparison_2021}.}

\section{Dark gap statistics \label{sec:dark_gap_stat}}
\subsection{Method} \label{sec:methods}
We define a dark gap to be a continuous spectral region in which all pixels binned to $1 \cmpch$ have
an observed normalized flux $F=F_{\rm obs}/F_{\rm c} < 0.05$, 
where $F_{\rm obs}$ is the observed flux and $F_{\rm c}$ is the continuum flux. The minimum length of a dark gap is $1\cmpch$.  We apply this definition when searching for dark gaps in both the real data and the mock spectra.  
The bin size and flux threshold were chosen to enable a uniform analysis over our large sample of spectra. 
A bin size of $1 \cmpch$ (corresponding to a velocity interval of $\Delta v\simeq 150~{\rm km\,s^{-1}}$ at $z=5.6$) provides a convenient scale that preserves most of the structure of the \lya\ forest.  
The choice of the flux threshold $F_{\rm t}$ is mainly restricted by the quality
of the data. 
Our choice of $F_{\rm t} = 0.05$ corresponds to non-detection of transmission lower than approximately twice the binned flux error ($2\sigma$) in the spectrum with the lowest S/N in our sample.
Using such a threshold, all dark gaps longer than \add{$30\cmpch$} have $\tau_{\rm eff} > 4$. 
\footnote{Throughout this paper, $\tau_{\rm eff}$ of a dark gap is calculated based on flux averaged along the full length of the gap rather than over windows of a fixed length. \add{Most low $\teff$ values for short dark gaps are caused by skyline subtraction or telluric correction residuals.}}
We have tested that using 0.1 or 0.025 for the flux threshold does not change our conclusions fundamentally when applying the same criteria to both the observed and mock spectra.  \add{Setting $F_{\rm t} = 0.1$ tends to yield dark gaps that are less opaque, while setting $F_{\rm t} = 0.025$ would decrease the number of usable QSO sightlines from 55 to 37.}

In order to avoid the QSO proximity region, we identify dark gaps in the \lya\ forest starting from
7 proper Mpc (pMpc) blueward from the QSO, which is close to the size of the largest proximity zones of bright QSOs at these redshifts  \citep{eilers_implications_2017,eilers_detecting_2020}. 
On the blue end, we limit our search to greater than 1041 \AA\ in the rest frame in order to avoid contamination from associated \lyb\ or \ion{O}{6} absorption \citep[e.g.,][]{becker_evidence_2015}.  For the purpose of comparing our results to simulations, we wish to avoid dark gaps that may be truncated by transmission peaks within the proximity zone. 
When quantifying the fraction of lines of sight that intersect gaps of length $L \ge 30 \cmpch$ (Sections \ref{sec:F30} and \ref{sec:CDF}), the highest redshift at which we register an individual sightline that shows a long gap, \textit{if any}, is therefore $30 \cmpch$ blueward of our proximity zone cut, although the gap may include pixels that extend up to the proximity zone. 
Nevertheless, we still record the full lengths of gaps extending to this $30 \cmpch$ ``buffer zone'' when searching for the longest possible dark gaps in both data and simulations. Dark gaps completely located in the QSO proximity zone and/or in this ``buffer zone'', however, are discarded. This ensures that the \add{pixel at} the red end of each sightline \add{\textit{may}} intersect a long ($L \ge 30 \cmpch$) dark gap.\footnote{If we do not introduce this ``buffer zone'', there is a possibility that the $F_{30}$ (Section \ref{sec:F30}) is underestimated near the red end of a sightline, since there can exist otherwise $>30\cmpch$ gaps that are truncated by the edge or peaks in the proximity zone.}
Finally, we limit our analysis to $z < 6$ because the mean transmitted flux at $z>6$ is so low that 
\add{most spectra show long dark gaps, making the dark gap statistics less informative.}

We note that there is no perfect way to handle the proximity zone effect. 
It is difficult to precisely define and measure the proximity zone size for each QSO, which partly motivates our choice to use a fixed proximity zone cut. 
The proximity zone for the brightest QSOs in our sample (e.g., SDSS J0100+2802 and VDES J0224+4711) may be larger than our adopted cut of 7 pMpc.
Fortunately, the use of an \add{additional} $30 \cmpch$ buffer zone minimizes the potential effect of the larger proximity zone of these objects.
In addition, because we limit our statistics over $5<z<6$, proximity zone transmission at $z>6$ towards some extremely bright QSOs does not impact our results.
Still, one should treat dark gaps near the QSO proximity zone with caution.

Noisy residuals from skyline subtraction and telluric correction may divide an otherwise continuous region of depressed flux. To deal with this, when searching for dark gaps we mask out $\pm 75\,{\rm km\,s^{-1}}$ intervals of the spectra centered at peaks in the flux error array, which typically correspond to skyline residuals. The exception to this is that we do not mask out any pixels with $F>3\sigma_F$. For consistency, we apply the same masking procedure to the mock spectra. 
\footnote{Since we add noise to the mock spectra pixel-wise according to the noise array of each observed spectrum with a \add{Gaussian} distribution, the skyline residuals in the mock spectra are not actually modeled. However, masking $\pm 75\,{\rm km\,s^{-1}}$ intervals makes the profile of sky subtraction residuals unimportant.}
In Appendix \ref{app:no_masking}, we use the mock spectra to show that such masking only produces a minor change in the results. We also test that the impact of masking telluric correction residuals near 7600-7650 \AA\ is neglectable.

As for contamination from damped \lya\ systems (DLAs) or other metal-enriched absorbers, we made no correction for their effect on dark gap \add{detection} following, e.g., \citet{fan_constraining_2006}.  Even strong DLAs can hardly, on their own, produce dark gaps as long as 30 $\cmpch$, which are the primary focus of this work. Nevertheless, in the results we label \add{dark gaps with intervening metal systems} for reference based on the systems identified by \cite{chen_mg_2017} and \cite{becker_evolution_2019}, as well as our own inspections. We visually searched all $L \ge 30 \cmpch$ dark gaps for metal absorbers not listed in the literature. The systems were identified via the coincidence of multiple metal lines in redshift. The metal lines we used include \ion{C}{2} $\lambda 1334$, \ion{C}{4} $\lambda\lambda 1548,1550$, \ion{O}{1} $\lambda 1302$, \ion{Mg}{2} $\lambda \lambda 2797,2803$, \ion{Al}{2} $\lambda 1670$, \ion{Si}{2} $\lambda 1527$, and \ion{Si}{4} $\lambda\lambda 1394,1403$.  A detection required these metal lines (if available) to have significant absorption features and self-consistent velocity profiles at the same redshift.  We have a good wavelength coverage for most metal lines mentioned above in QSO spectra taken with X-Shooter.  Even for these objects, however, we caution that the list of metal absorbers may be still incomplete.  A full list of metal absorbers in the XQR-30 spectra will be presented by R. Davies et al., in prep.
We also note that the simulations we used do not include DLAs or other metal-enriched absorbers.

\subsection{Notable dark gaps}

\begin{figure*}[ht]
    \begin{center}
        \includegraphics[width=6.in]{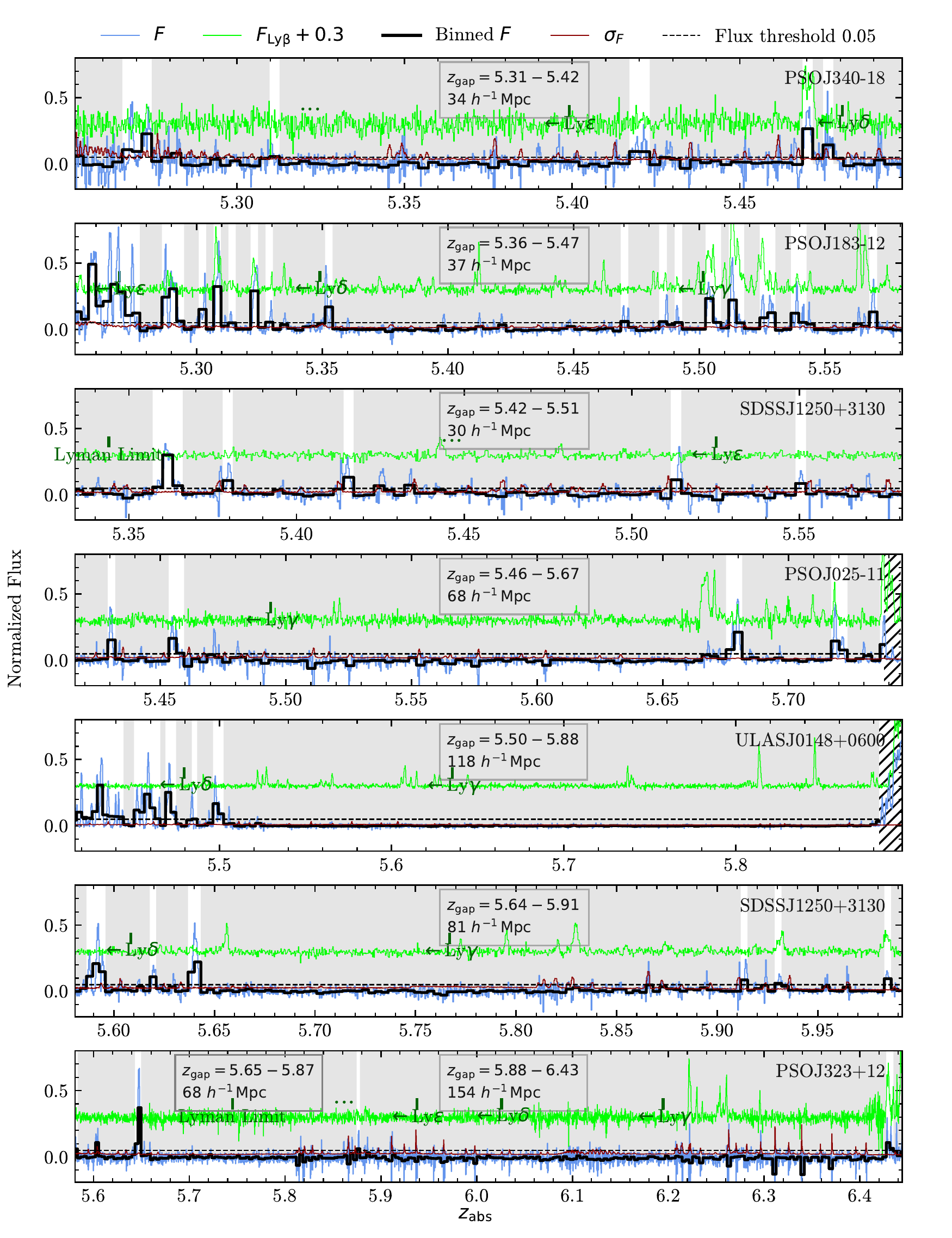}
    \end{center}
    \caption{Examples of notable dark gaps identified. Dark gaps are labeled with gray shades. The thick black line is the binned flux with binning size of $1\cmpch$. Dashed black horizontal line sets the flux threshold of 0.05. Un-binned flux and flux uncertainty are represented by thin blue and dark red lines, respectively. The text boxes display the redshift span and length of each long dark gap ($L \ge 30\cmpch$). Regions redward the proximity zone cut are labeled with hatches and are excluded from the statistics. For reference, the green line, offset by 0.3 in flux, is shifted in wavelength to show the \lyb\ and higher-order Lyman forest at the same redshifts. Vertical ticks label the starting points of higher-order Lyman forests.  
    \label{fig:exmaple_gaps}}
\end{figure*}

Long dark gaps play an important role in characterizing the IGM in the later stages of reionization. 
Among \add{50} dark gaps with $L \ge 30\cmpch$ detected in our sample, Figure \ref{fig:exmaple_gaps} displays some notable examples. They either extend down to or below $z\sim5.5$, are extremely long ($L>80\cmpch$), or both.

Two long dark gaps \add{entirely} at $z<5.5$ are identified towards PSO J183-12 and PSOJ340-18. They span $z_{\rm gap} = 5.36-5.47$ and $z_{\rm gap} = 5.31-5.42$, corresponding to lengths of $L=37 \cmpch$ and $L=34\cmpch$, respectively. Most spikes and sharp dips with negative flux in the un-binned spectra inside the two gaps are skyline subtraction residuals as indicated by the peaks in the flux error array. Both dark gaps are highly opaque, with $\teff > 6$. The spectra of both QSOs have a good coverage of redshifted common metal lines. We searched their X-Shooter VIS and NIR spectra and found no metal absorption within the redshift ranges of the dark gaps. \add{In addition, a $30\cmpch$ dark gap extending just above $z=5.5$ is identified towards SDSS J1250+3130. Most of the spikes inside this gap are also probably due to sky lines as indicated by peaks in the flux error array.}

The \add{fourth through sixth} rows in Figure \ref{fig:exmaple_gaps} display three examples of long dark gaps extending down to $z\sim5.5$. The long gap extending to $z=5.46$ with a length of $L=68 \cmpch$ towards PSOJ025-11 is one of the longest troughs below redshift six discovered in this work. The only weak transmission peaks in the un-binned flux array that seem to be real are the ones at $\zabs\simeq5.47$, 5.48, and 5.67. Overall, however, it is extremely dark, with $\teff\geq 6.4$. We also reproduce the detection of the long trough discovered towards ULAS J0148+0600 by \citet{becker_evidence_2015}, which extends down to $z=5.5$ with a total length $L>110\cmpch$. Due to the use of a different definition of dark gap compared to \citet{becker_evidence_2015}, the trough detected in this work includes an additional small transmission peak that appears in the un-binned spectrum near the blue end.  This yields a slightly larger $L$ but a comparable $\teff$ value. We also find a gap of $L=81 \cmpch$ extending down to $z=5.64$ towards SDSS J1250+3130. 
Spikes within the trough are skyline subtraction residuals, 
as shown by peaks in the error array. We do not see any strong metal absorbers that would indicate dense absorption systems such as DLAs or Lyman limit systems (LLSs), in any of these gaps. 
Finally, we find dark gaps longer than $110\cmpch$ towards several QSOs
with the highest redshifts in our sample. This is not surprising because the IGM is more neutral at higher redshifts and therefore more likely to produce large
\lya\ opaque regions. For example, \citet{barnett_observations_2017} identified a 240$\cmpch$ gap at $z > 6.1$ towards the $z = 7.1$ QSO ULAS J1120+0641.  Here we display a remarkably long dark gap towards PSO J323+12. It covers $z_{\rm gap}=5.88$--6.43 and has a length of $154 \cmpch$, as shown in the bottom row of Figure \ref{fig:exmaple_gaps}. 

For reference, we overplot in green the regions of spectra corresponding to \lyb\ for the \lya\ shown in Figure~\ref{fig:exmaple_gaps}.\footnote{We use a power-law fit to the continuum for regions blueward of the \lya\ forest because our PCA implementation does not cover these wavelengths; however, this should not significantly affect the qualitative results for the higher-order Lyman series transmission shown in Figure~\ref{fig:exmaple_gaps}.}  In many cases the \lyb\ forest also includes higher order Lyman series absorption, as indicated in the figure.  Although dark gaps are highly opaque to \lya, there are often narrow transmission peaks corresponding to \lyb.  These peaks demonstrate that the dark gaps in \lya\ typically cannot arise from continuous regions of neutral gas, which would be highly opaque to all Lyman series lines.  Broken regions of neutral gas may still be present, however, with the \lyb\ transmission corresponding to gaps between neutral sections \citep[e.g.,][]{keating_long_2020,nasir_observing_2020}.

\subsection{Overview of dark gaps}\label{sec:overview}
In total, we detected \add{1329} dark gaps from the sample, of which \add{50} have a length of $L \ge 30 \cmpch$. Properties of all dark gaps detected are summarized in Table \ref{tab:dark gaps}. Details on dark gap detection for each QSO sightline are shown in Figure Set \ref{fig:appendix01}.

\begin{deluxetable*}{rlllcrc}
    \tablenum{2}
    \tablecaption{Properties of dark gaps \label{tab:dark gaps}}
    \tablehead{
        \colhead{Index} & 
        \colhead{QSO} & \colhead{$z_{\rm blue}$} & \colhead{$z_{\rm red}$} &
        \colhead{$L~(\cmpch)$} & \colhead{$\teff$} & \colhead{$z_{\rm absorber}$}
    }
    \decimalcolnumbers
    \startdata
    26 & ULASJ1319+0950 & 5.876 & $6.012^{\rm b}$ & $\geq 40 $ & $\geq 6.674$ &  \\
    \add{157} & SDSSJ0100+2802 & 5.883 & 5.988 & $31 $ & $7.874 \pm 0.339$ & 5.945, 5.940 \\
    \add{240} & PSOJ108+08 & 5.661 & $5.836^{\rm b}$ & $\geq 54 $ & $6.062 \pm 0.145$ &  \\
    281 & PSOJ183-12 & 5.332 & 5.350 & $6  $ & $4.193 \pm 0.133$ &  \\
    292 & PSOJ183-12 & 5.690 & 5.702 & $4  ^{\rm c}$ & $3.951 \pm 0.082$ &  \\
    294 & SDSSJ1602+4228 & 5.065 & 5.071 & $2  $ & $3.580 \pm 0.251$ &  \\
    350 & ATLASJ025.6821-33.4627 & $5.285^{\rm a}$ & 5.356 & $24 $ & $5.680 \pm 0.226$ &  \\
    \add{817} & SDSSJ1148+5251 & 5.853 & $6.285^{\rm b}$ & $\geq 124$ & $\geq 7.558$ & 6.258, 6.011, 6.131 \\
    \add{959} & SDSSJ1137+3549 & 5.683 & 5.686 & $1  $ & $\geq 3.569$ &  \\
    \enddata
    \tablecomments{Columns: 
    (1) index of the dark gap,
    (2) QSO name,
    (3) redshift at the blue end of the gap,
    (4) redshift at the red end of the gap,
    (5) dark gap length,
    (6) effective opacity of the dark gap based on the flux and flux error in the original binning,
    (7) redshift(s) of known metal absorber(s) in the dark gap, if any. \\
    $^{\rm a}$ Dark gap starting at the blue edge of the \lya\ forest. \\
    $^{\rm b}$ Dark gap ending at the red edge of the \lya\ forest.    \\
    $^{\rm c}$ Dark gap located completely inside the buffer zone.\\
    (This table is published in its entirety in the machine-readable format.
    A portion is shown here for guidance regarding its form and content.)
    }
\end{deluxetable*}

\begin{figure}[htb]
            \centering
    \includegraphics[width=3.3in]{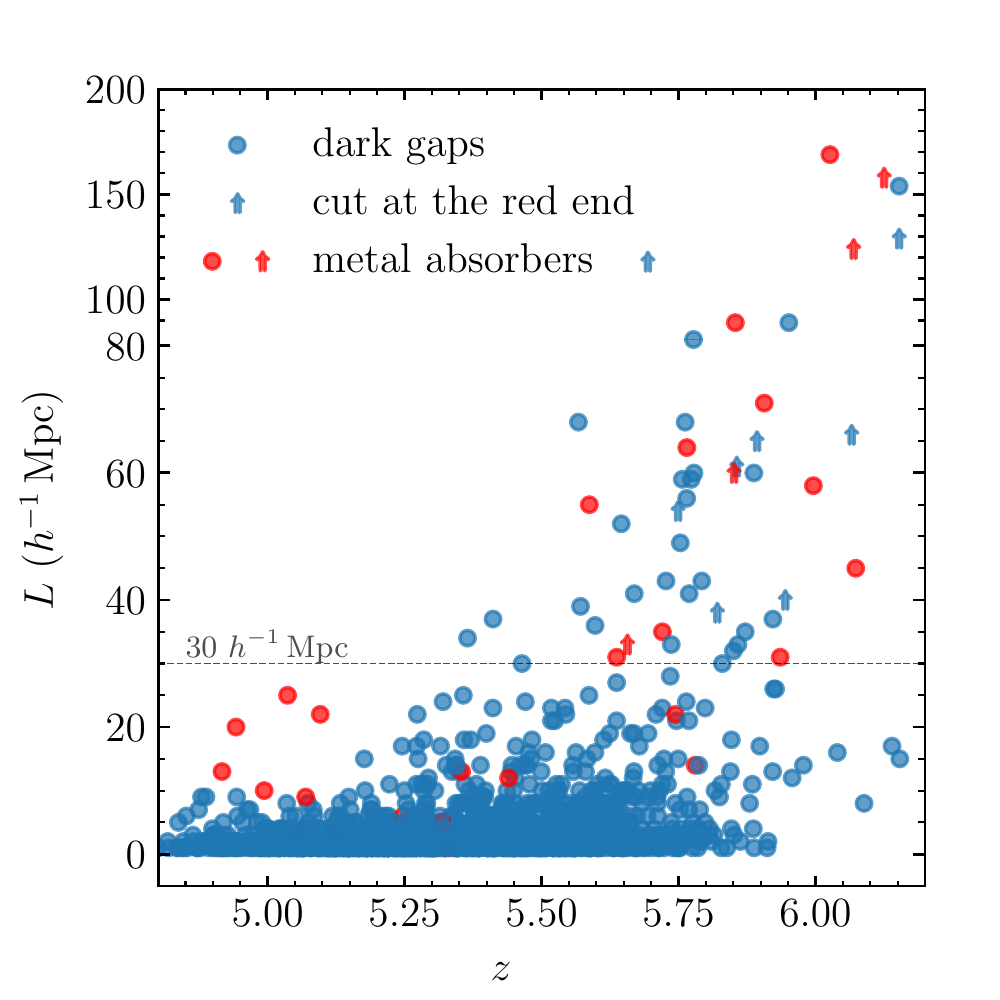}
    \caption{Gap length versus central redshift
    for dark gaps detected from our sample.
    Dark gaps located completely in the ``buffer zone'' are excluded from this plot.
    Arrows indicate dark gaps whose red edge lies within $7~{\rm pMpc~}$ from the QSO and are therefore potentially truncated by the proximity effect; lower limits on the length are therefore given for these gaps.  
    Red symbols indicate dark gaps with one or more metal absorbers.
    \label{fig:scatter}}
\end{figure}

\begin{figure*}[ht]
    \begin{center}
        \includegraphics[width=6.2in]{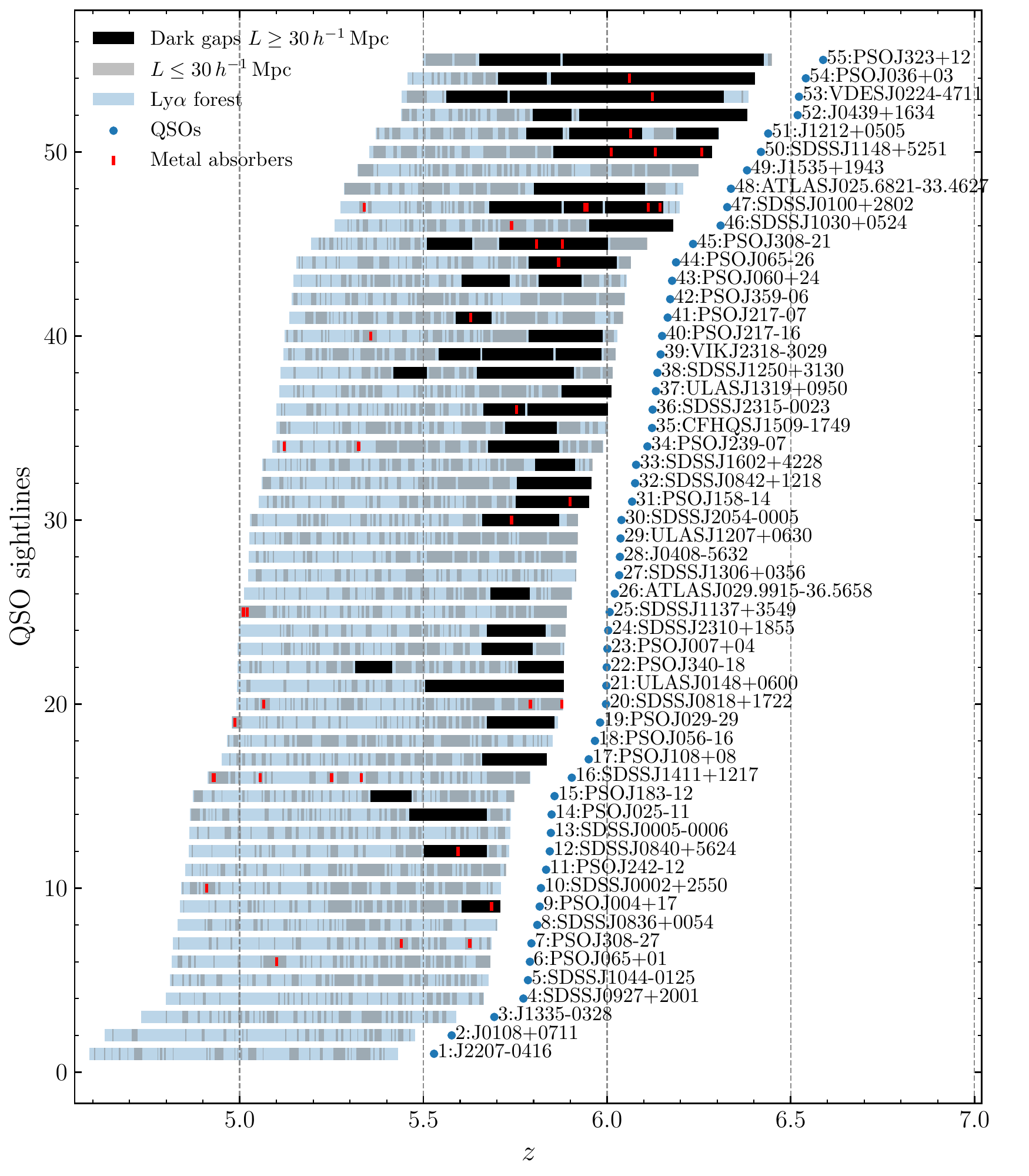}
    \end{center}
    \caption{Overview of dark gaps identified in the \lya\ forest of 55 QSO sightlines.
    Black bars and gray shaded regions represent dark gaps longer and shorter than $30\cmpch$, respectively. Red short lines denote known associated metal absorbers intervening dark gaps. Light blue shaded regions indicate the redshift coverage of the \lya\ forest. Redshifts of QSOs are marked with blue dots. The Ly$\alpha$ forest is truncated at 7 pMpc from the QSO. The \lya\ forest shown in this figure \textit{includes} the $30 \cmpch$ buffer zone on the red end, which is excluded from the statistical analysis of dark gaps. See Section \ref{sec:methods} for details.
        }
    \label{fig:all_los}
\end{figure*}

As an overview, Figure \ref{fig:scatter} plots all dark gaps identified in this work according to their central redshift and length. Dark gaps with associated metal absorbers are labeled in red. This figure has excluded dark gaps that are completely inside the 7 proper-Mpc proximity zone and/or inside the $30\cmpch$ ``buffer zone'' beyond the proximity zone. Not surprisingly, as redshift increases, there are more long dark gaps and a larger scatter in dark gap length. The lowest-redshift gaps with $L \ge 30\cmpch$ appear around $z=5.3$. 

Figure \ref{fig:all_los} displays the \lya\ forest 
coverage and all dark gaps identified for every line of sight in our sample. At $z\lesssim5.2$, most QSO sightlines are highly transmissive; a few gaps with $L \sim 10$--20$\cmpch$ appear but these tend to contain metal absorbers and are likely to be DLAs.
Dark gaps longer than $30 \cmpch$ appear in the sightlines of PSO J340-18 and PSO J183-12 at $z\simeq 5.3$ and 5.4, respectively.
The frequency of long dark gaps increases with redshift such that most lines of sight at $z \simeq 5.8$ show gaps longer than 30$\cmpch$ in the \lya\ forest.
Interestingly, the J1535+1943 sightline is relatively transmissive at $z\sim6$ compared to others at the same redshift. Although J1535 has a reddened spectrum, the continuum re-construction is acceptable and most of the transmission peaks in the \lya\ forest appear to be real.

\subsection{Fraction of QSO spectra exhibiting long dark gaps}\label{sec:F30}
We introduce the fraction of QSO spectra exhibiting long ($L \ge 30 \cmpch$) gaps as a function of redshift, $F_{30}(z)$, as a new \lya\ forest statistic. \add{As mentioned in Section \ref{sec:methods}, in order to deal with the finite length of the spectra for this statistic we cut off each QSO sightline at the blue edge of the $30\cmpch$ buffer zone.} $F_{30}$ quantifies how common the large Lyman-alpha-opaque regions are and how they evolve with redshift.  
We choose $30 \cmpch$ because
we found that this length most effectively distinguishes between the models described in Section \ref{sec:simulations}, especially between the {\tt homogeneous-UVB} and other models.
The comparison of \add{the dark gap length distribution,} $P(L)$\add{,} predicted by different models in Section \ref{sec:NL_PDF} also implies that dark gaps with $L \ge 30\cmpch$ are potentially good probes for \ion{H}{1} if the late reionization scenario is indeed preferred. We note that we include all long dark gaps regardless of the presence of associated metal absorbers since the dense absorption systems alone are not likely to create troughs longer than $30\cmpch$.

\begin{figure}
    \begin{center}
        \includegraphics[width=3.355in]{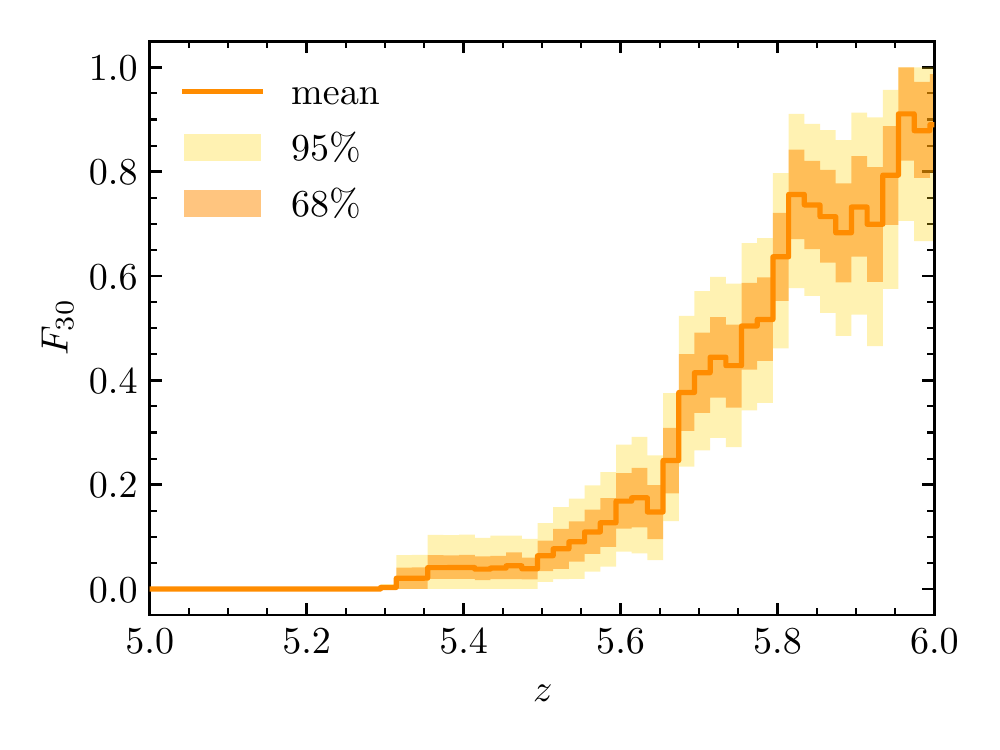}
        \caption{Measured fraction of QSO spectra exhibiting long ($L \ge 30 \cmpch$) dark gaps as a function of redshift.
                We use bootstrap re-sampling to calculate the mean, 68\% and 95\% limits of $F_{30}$ averaged over $\Delta z = 0.02$ bins, presented with the solid orange line, dark and light shaded regions, respectively. \\(Tabular Data behind the Figure (DbF) is available online.)}
        \label{fig:F30-obs}
    \end{center}
\end{figure}

Figure \ref{fig:F30-obs} displays the evolution of $F_{30}$ with redshift measured from the QSO spectra. The result is averaged over $\Delta z = 0.02$ bins. The mean, 68\% limits, and 95\% limits of $F_{30}$ are calculated based on 10000 bootstrap re-samplings of the whole sample. In each realization, we randomly select 55 QSO spectra, with replacement, and add up the number of sightlines \add{showing} $L \ge 30\cmpch$ dark gaps at a given redshift. The total is then normalized by the number of QSO sightlines at each redshift, which yields $F_{30}(z)$ for this realization. 
$F_{30}$ starts to be nonzero from $z\simeq5.3$ and increases strongly with redshift.  At $z=6$, $\sim$90\% \add{of} sightlines \add{present} long gaps.

We noted above that a deep spectrum of SDSS J1250+3130 was obtained based on preliminary indications from shallower data of a long gap in its spectrum.  
This is the only QSO in the sample for which
the selection 
is related to the foreknowledge of dark gaps.  We include J1250 for completeness, but note that excluding this line of sight from our sample would only decrease (increase) $F_{30}$ by $\lesssim0.02$ (0.05) over $5.50<z<5.90$ ($5.90<z<5.93$).

\add{Finally, we test whether metal absorbers could be linking adjacent dark gaps in a way that would impact our $F_{30}$ statistic.  For this we calculate a ``pessimistic'' $F_{30}$ by dividing dark gaps at the redshifts of DLAs and other metal systems (Appendix \ref{app:nodla}). The resulting change in $F_{30}$ is minor, with a maximum decrease of $\sim0.1$ at $z\sim5.8$.  The differences between the observations and model predictions (Section \ref{sec:model_comparison}; Figure \ref{fig:F30nodla}) can still be well distinguished.  We therefore conclude that this potential impact of metal absorbers on $F_{30}$ is not significant.}

\subsection{Distribution of dark gap length \label{sec:CDF}}

\begin{figure}[ht]
    \centering
    \includegraphics[width=3.355in]{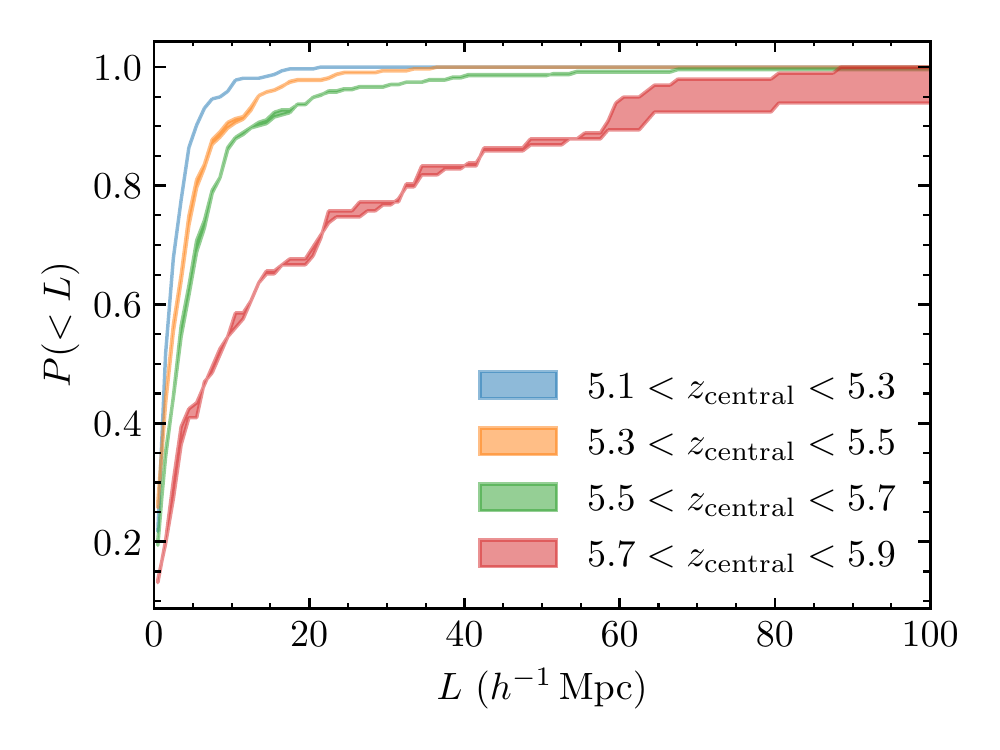}
    \caption{Cumulative distributions of dark gap length. We include all dark gaps regardless of the presence of associated metal absorbers. The upper and lower bounds of the shaded regions correspond to the most pessimistic and optimistic cases for $P(<L)$. See text for details.
    \label{fig:CDF_obs_cent}}
\end{figure}

In addition to $F_{30}$, we investigate the cumulative distribution function (CDF) of gap length, $P(<L)$.
Figure \ref{fig:CDF_obs_cent} plots $P(<L)$ in redshift bins of $\Delta z = 0.2$.  Dark gaps are assigned to a bin based on the central redshift of the gap, and we do not truncate gaps extending beyond the edges of the redshift windows. We treat the dark gaps truncated by the 7 proper-Mpc proximity zone cut by plotting the most pessimistic and optimistic bounds on $P(<L)$. The pessimistic bound is calculated by considering the lengths of dark gaps are as measured. The optimistic bound, however, is given by assuming the lengths of truncated dark gaps are infinite, which indicates the most extreme dark gap length possible in the absence of the QSO. In the latter case, we still use the measured central redshift of each dark gap to assign it to a redshift bin. 

Figure \ref{fig:CDF_obs_cent} demonstrates that longer dark gaps become more common towards higher redshifts. This is consistent with the result of $F_{30}$. Moreover, similar to the rapid redshift evolution in $F_{30}$ near $z\simeq5.7$, $P(<L)$ shows a large change between $5.5<z<5.7$ and $5.7<z<5.9$.

To test the effects of metal absorbers on $P(<L)$, we calculate the distribution by excluding dark gaps with known associated metal absorbers. We find the difference is minor. The maximum increment on the most pessimistic $P(<L)$ over $5.7<z<5.9$ is less than 0.03, and the difference is less than 0.005 over the other redshift bins.

\section{Models and Simulations for Comparison\label{sec:simulations}}

\begin{figure}
    \centering
    \includegraphics[width=3.5in]{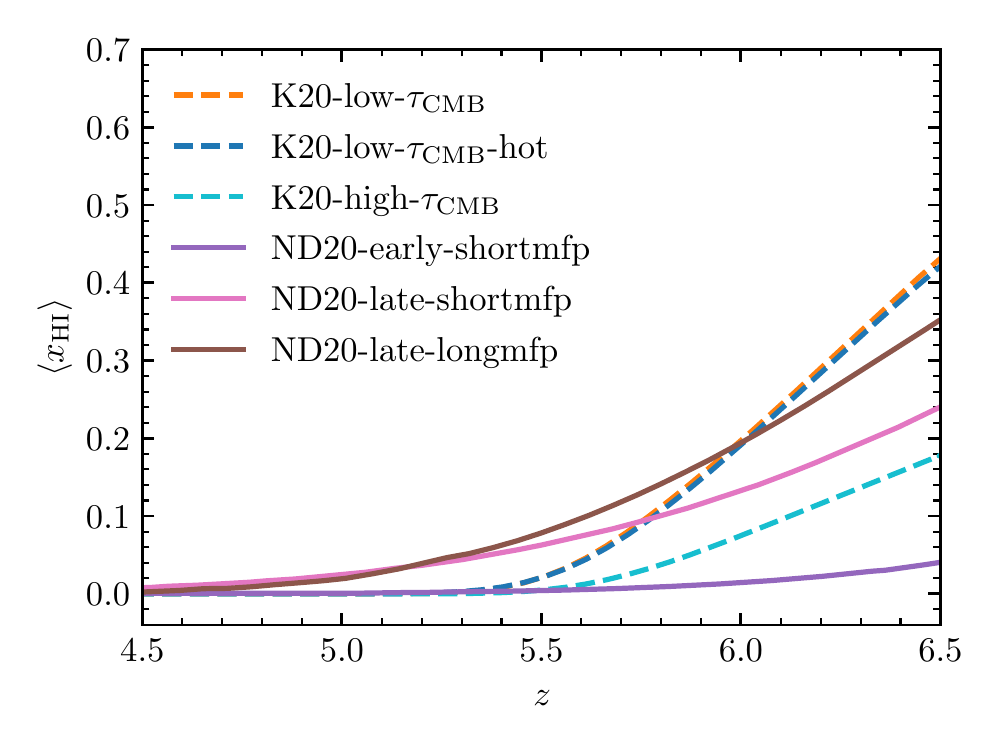}
    \caption{Redshift evolution of the volume-weighted average neutral hydrogen fraction of reionization models used in this work. See Table \ref{tab:models} for the key information of these models and details in Section \ref{sec:simulations}.}
    \label{fig:neutral_fraction}
\end{figure}

We compare our measurements to predictions from hydrodynamical simulations 
that span a range of reionization histories and UV backgrounds.  
Here we briefly describe the simulations.  
The key information is summarized in Table \ref{tab:models}, with the redshift evolution of the volume-weighted neutral hydrogen fraction $\langle x_{\rm HI} \rangle$ for each simulation plotted in Figure \ref{fig:neutral_fraction}.

\begin{deluxetable}{lccc}
    \tablenum{3}
    \tablecaption{Models used in this work \label{tab:models}}
    \tablehead{
        \colhead{Model} & 
        \colhead{Reionization} &
        \colhead{$~~z_{95}~~$} & 
        \colhead{$~~z_{50}~~$}
    }
    \decimalcolnumbers
    \startdata
        {\tt homogeneous-UVB}              & -      & 15      &  -   \\
        \hline                                      
        {\tt K20-low-$\tau_{\rm CMB}$}     & late   &  5.6   & 6.7  \\
        {\tt K20-low-$\tau_{\rm CMB}$-hot} & late   &  5.6   & 6.7  \\
        {\tt K20-high-$\tau_{\rm CMB}$}    & late   &  5.9   & 8.4  \\
        {\tt ND20-late-longmfp}            & late   &  5.3   & 7.0  \\
        {\tt ND20-late-shortmfp}           & late   &  5.4   & 7.5  \\
        \hline                                          
        {\tt ND20-early-shortmfp~~~~~~~}   & early  &  6.6   & 8.7  \\
    \enddata
    \tablecomments{Columns: (1) name of the model,
    (2) qualitative description of the reionization model,
            (3) redshift at which the volume filling factor of ionized gas reaches 95\%,
    (4) redshift at which the volume filling factor of ionized gas reaches 50\%,
    We use {\tt K20} for models from \citet{keating_constraining_2020} and {\tt ND20} for models from 
    \citet{nasir_observing_2020}. See Sections \ref{sec:late_reionization} \& \ref{sec:early_reionization} for details.
    }
\end{deluxetable}

\subsection{Homogeneous UV Background}
We first include a baseline model wherein reionization is fully completed 
at $z > 6$ and the UVB is spatially uniform.  
For this we use a run from the Sherwood simulation suite, 
which successfully reproduces multiple characteristics of the observed
\lya\ forest over $2.5 <z<5$ \citep{bolton_sherwood_2017}.  
The Sherwood suite uses a homogeneous 
\citet{haardt_radiative_2012} UV background.  
Reionization occurs instantaneously at $z = 15$, allowing the IGM to fully 
relax hydrodynamically by $z = 6$. 
The simulations were run with the parallel smoothed particle hydrodynamics 
code {\tt P-GADGET-3},
which is an updated and extended version of
{\tt GADGET-2} \citep{springel_cosmological_2005}.
We use the simulation with $2\times 2048^3$ particles and box size of
$(160 \cmpch)^3$ to build mock spectra for the {\tt homogeneous-UVB} model, as described in Section \ref{sec:mock}. 

\subsection{Late Reionization \label{sec:late_reionization}}
We use two sets of models wherein reionization continues significantly below redshift six.  In these models, long dark gaps in \lya\ transmission at $z < 6$ arise from a combination of neutral islands and regions of suppressed UVB, which are often adjacent to one another.

The first late reionization models are from \citet{keating_constraining_2020}. They include three models with different ionization and/or thermal histories. We denote the fiducial model as {\tt K20-low-$\tau_{\rm CMB}$}, wherein the volume filling fraction of ionized gas reaches 95\% at $z=5.6$ and 99.9\% at $z=5.2$.
Two other runs, the {\tt K20-low-$\tau_{\rm CMB}$-hot} and 
{\tt K20-high-$\tau_{\rm CMB}$} models, are also included.
Briefly, the {\tt K20-low-$\tau_{\rm CMB}$-hot}
model uses a higher temperature for the input blackbody ionizing spectrum, namely $T=40000~{\rm K}$ instead of $T=30000~{\rm K}$ as used in the {\tt K20-low-$\tau_{\rm CMB}$} model. They have a volume-weighted mean temperature at the mean density at $z = 6$ of $T_0\simeq10000$ K and 7000 K, respectively.
The {\tt K20-high-$\tau_{\rm CMB}$} model shares a similar IGM thermal history
with the {\tt K20-low-$\tau_{\rm CMB}$} model, but it has an earlier reionization midpoint of 
$z_{\rm mid}=8.4$.

The {\tt K20} simulations are modified 
versions of the late reionization model published in \citet{kulkarni_large_2019}.
The model was modified such that
(i) the IGM temperature evolution is in better agreement with recent observations \citep{boera_revealing_2019,walther_new_2019,gaikwad_consistent_2020},
and (ii) the mean \lya\ transmission is in better agreement with data at $z < 4.7$ \citep[][]{becker_evidence_2015}.
The ionization state of the IGM is modeled using the radiative transfer code {\tt ATON} \citep[][]{aubert_radiative_2008,aubert_reionization_2010} that post-processes underlying hydrodynamic simulations performed with {\tt P-GADGET-3}.
The simulations use the identical initial condition and box size of the Sherwood simulation suite. The radiative transfer, however, leads to an extended and self-consistent reionization history. This produces scatter in the \lya\ $\teff$. The simulations also contain fluctuations in temperature and photoionization rates. A lightcone from the radiative transfer simulation were extracted on the fly. Using sightlines through this lightcone, \citet{keating_constraining_2020} computed the optical depths continuously spanning $4.0 \lesssim z \lesssim 7.5$ for each model, which allows us to avoid having to do any interpolation.

The second set of late reionization models is from \citet{nasir_observing_2020}. In these models, the volume filling factor of ionized gas reaches 95\% at $z = 5.3$--5.4.  As in the \citet{keating_constraining_2020} models, fluctuations in both the UVB and temperature are present. The UVB fluctuations are driven by a short and spatially variable mean free path, similar to the model in \citet{davies_large_2016}.  In the two \citet{nasir_observing_2020} models, which we denote as {\tt ND20-late-longmfp} and {\tt ND20-late-shortmfp}, the volume-weighted average mean free path for 
$912$ \AA\ photons at $z = 5.6$ is $\langle \lambda_{\rm mfp}^{912} \rangle = 30 \cmpch$ and $10 \cmpch$, respectively. As a result of the shorter mean free path, {\tt ND20-late-shortmfp} contains stronger fluctuations in the UVB. The shorter $\langle \lambda_{\rm mfp}^{912} \rangle$ is also more consistent with the recent mean free path measurement of \citet{becker_mean_2021}.

The \citet{nasir_observing_2020} simulations use a modified version of the Eulerian hydrodynamics code from \citet{trac_moving_2004}. 
They use $2\times 2048^3$ gas and dark matter resolution elements and a box size of $L = 200 \cmpch$. 
To model the effects of reionization on the forest, they post-process the hydrodynamics simulations using semi-numeric methods. Optical depth skewers are available at $z=5.6$, 5.8, and 6.0, and neutral fraction information is available at $z=5.6$ and 5.8.  A sample of 4000 lines of sight were extracted at each redshift, with each optical depth skewer having a length of $500\cmpch$ by making use of the periodic boundary conditions (F. Nasir, private communication).

\subsection{Early reionization with a fluctuating UVB \label{sec:early_reionization}}

Finally, we include a model from \citet{nasir_observing_2020} wherein the volume filling factor of ionized gas reaches $\sim98\%$ by $z = 6$ 
but the UVB retains large spatial fluctuations to somewhat lower redshifts.\footnote{The volume filling factor of ionized gas no longer increases significantly at $z<6$. Although it has not reached 99\% strictly by $z=6$, we still consider this model as an early reionization model.} 
It has $\langle \lambda_{\rm mfp}^{912} \rangle = 10 \cmpch$ as in the {\tt ND20-late-shortmfp} model. We refer to this model as {\tt ND20-early-shortmfp}. 
It is essentially a modified version of the fluctuating UVB model proposed by \citet{davies_large_2016} with temperature fluctuations included. Compared to the {\tt ND20-late-shortmfp} model mentioned previously, the {\tt ND20-early-shortmfp} model has a similarly broad UVB distribution but a much earlier end of reionization.
In this model, long dark gaps at $z < 6$ primarily correspond to regions with a low UVB. Since the IGM is not technically fully ionized in this model until down to $z\simeq5$, however, a small fraction of dark gaps may still contain some neutral hydrogen.

\subsection{Construction of mock spectra \label{sec:mock}}

In order to directly compare the observations to the models
we construct mock spectra from the simulations with properties similar to the real data. We firstly describe how we create mock spectra for the {\tt homogeneous-UVB} model.

The snapshots for the {\tt homogeneous-UVB} model are available on every $\Delta z=0.1$ interval over $3.9\leq z \leq 8.9$.  To be consistent with the simulations from  \citet{nasir_observing_2020} we only use snapshots from every $\Delta z = 0.2$, and the same snapshots are used for every sightline.
We have verified, however, that using snapshots spaced every $\Delta z = 0.1$ would not significantly impact our results.  Each snapshot was used to extract 5000 $160\cmpch$ skewers along which the native \lya\ optical depths have been calculated \citep[][]{bolton_sherwood_2017}. For a mock spectrum centered at redshift $z_0$ we combine skewers from redshifts $z_0 - 0.2$, $z_0$, and $z_0 + 0.2$ \footnote{We cut the skewers into three pieces and then stitch the corresponding pieces with those from the adjacent redshifts. Only a portion of a $160\cmpch$ skewer from a given snapshot is therefore used for a mock spectrum centered at $z_0$.}
after shifting the periodic lines of sight by random amounts. \add{The resulting mock spectra are still $160\cmpch$ in length but contain information about the redshift evolution of the \lya-opaque regions.} We fit the $\teff$ evolution over $5\leq z\leq6$ from \citet{bosman_new_2018} with a power law of $\teff \propto (1+z)^{12.34}$ and re-normalize the optical depths of the mock spectra such that their average \lya\ transmission matches this evolution.
We have also checked that the mean transmission measured directly from our observed sample is within the $1\sigma$ uncertainties of the measurement in \citet{bosman_new_2018}.
  \add{We create 5000 mock spectra matching each of our 55 lines of sight. For each QSO,} we bin the mock spectra using exactly the same wavelength array  \add{as the observed} spectrum.  \add{We then} add Gaussian noise to the mock spectra based on  \add{the} corresponding  flux error array. 

Because each optical depth skewer \add{from the} \citet{nasir_observing_2020} models has a length of $500\cmpch$, we first clip them to $160\cmpch$ and then follow a similar procedure to build the mock spectra set at $z_0=5.8$ as described above, \add{including rescaling the effective optical depth}. In order to cover the full redshift range of the simulation outputs, we extend the mock spectra down to $z=5.6$ and up to $z=6.0$ by making use of the unclipped skewers to create mock spectra sets centered at $z_0=5.6$ and 6.0. However, since the spatial structure of the IGM is only recovered over $5.6\leq z \leq 6.0$, we \add{restrict} our dark gap analysis to this redshift range.
As for {\tt K20} models, \citet{keating_constraining_2020} ran many radiative transfer simulations until converged on a reionization history that self-consistently reproduces the mean flux of the \lya\ forest as measured by \citet{bosman_new_2018}.
We therefore only needed to re-bin the skewers and add noise in order to match them to each individual observed QSO spectrum. \add{We note that continuum errors are not considered for the mock spectra.  This is because the continuum errors for the observed spectra are estimated to be small ($\lesssim$10\%; Section~\ref{sec:fitting}), and partly because we are primarily concerned with very low flux levels, which are less affected in an absolute sense by continuum uncertainties.}

\begin{figure*}[ht]
    \centering
    \includegraphics[width=5.7in]{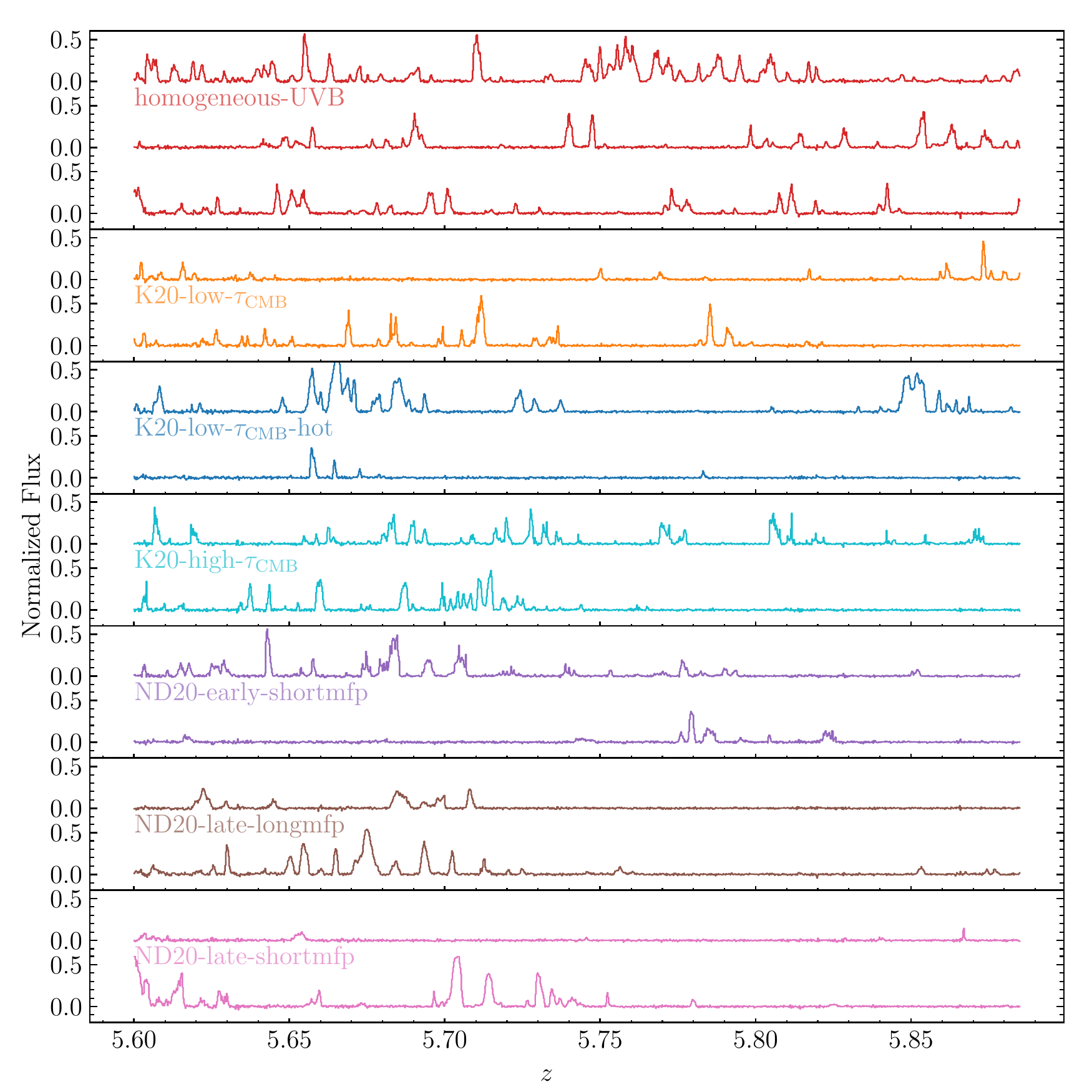}
    \caption{Example mock spectra with binning and S/N chosen to match the \lya\ forest of ULAS J0148+0600. 
    We randomly select mock sightlines from each model here. The colored lines represent the flux arrays.
    }
    \label{fig:mock}
\end{figure*}

In Figure \ref{fig:mock}, we display mock spectra randomly selected from all the models with S/N chosen to match the the \lya\ forest of ULAS J0148+0600 as examples. The {\tt homogeneous-UVB} model exhibits more small transmission peaks than the other models, as expected because the IGM is fully ionized by a uniform UVB.  The other models tend to show longer dark gaps interspersed with regions of high transmission.

\subsection{Neutral islands and dark gaps \label{sec:NL_PDF}}

Here we examine the connection between dark gaps and regions of neutral hydrogen.
For this we calculate the dark gap length distribution $P(L)$ predicted by models.
We use the method described in Section \ref{sec:methods} to find dark gaps in mock spectra generated in Section \ref{sec:mock}, but with no noise added, and identify gaps that contain regions of neutral hydrogen. The frequency of dark gaps with length $L$ for each model in each redshift bin is calculated based on 10000 realizations and normalized by the total count of dark gaps in each redshift bin, with $P(L)$ averaged over bins of $\Delta L = 5\cmpch$.
We consider a dark gap to contain neutral hydrogen if any pixels inside this gap have $x_{\rm HI} > 0.9$. Over each redshift bin, dark gaps extending beyond the boundaries of the $\Delta z = 0.2$ window are truncated at the edge. We do so to avoid artifacts in $P(L)$ caused by the finite length of the mock spectra.

\begin{figure*}[ht]
    \centering
    \includegraphics[width=6in]{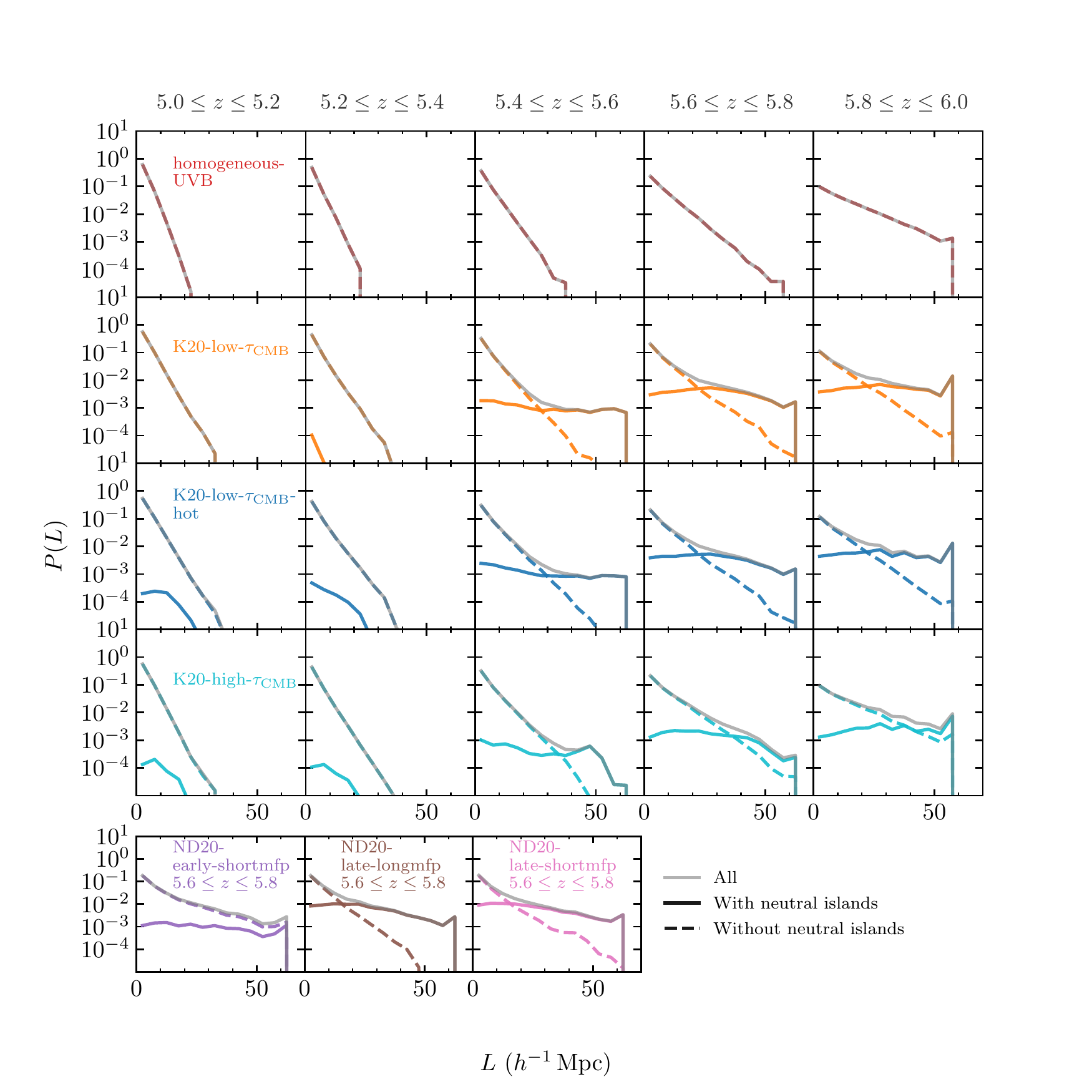}
    \caption{Distribution of dark gaps with and without neutral hydrogen predicted by simulations. A dark gap is considered to contain neutral gas if any pixels inside this gap have a neutral fraction of $x_{\rm HI} > 0.9$. $P(L)$ is calculated with count of dark gaps over $\Delta L = 5\cmpch$ bins divided by the total count of dark gaps. The distribution is calculated with dark gaps detected in 10000 sets of mock spectra (Section \ref{sec:mock}) for each simulation, but with no noise added. We note that the volume neutral fraction information of \citet{nasir_observing_2020} models is only available at $z=5.6$ and 5.8. The plots for {\tt ND20} models are therefore not extended to $z=6.0$.}
    \label{fig:neutral_NL}
\end{figure*}

As shown in Figure \ref{fig:neutral_NL}, $P(L)$ varies significantly between models. Firstly, no dark gaps with neutral pixels are found in the {\tt homogeneous-UVB} model because the IGM is fully ionized. In the {\tt ND20-early-shortmfp} model, the IGM is 98\% ionized by $z=6$, and therefore only a small fraction of dark gaps contain neutral islands. Dark gaps with no neutral islands also dominate in the {\tt K20-high-$\tau_{\rm CMB}$} model that has an extended reionization history. The situation is very different in the rapid late reionization scenarios, however. Dark gaps with neutral islands become dominant for $L \ge 15$--20$\cmpch$ in both {\tt ND20-late} models. Similarly, in the {\tt K20-low-$\tau_{\rm CMB}$(-hot)} model, dark gaps with neutral islands start to be the majority for $L\gtrsim 25$--30$\cmpch$ at $z>5.4$. Long dark gaps with $L \gtrsim 30\cmpch$ are therefore of potentially high interest in terms of identifying regions of the IGM that may contain neutral gas. This paper is therefore largely focused on these long gaps. 

\begin{figure*}[ht]
    \centering
    \includegraphics[width=6in]{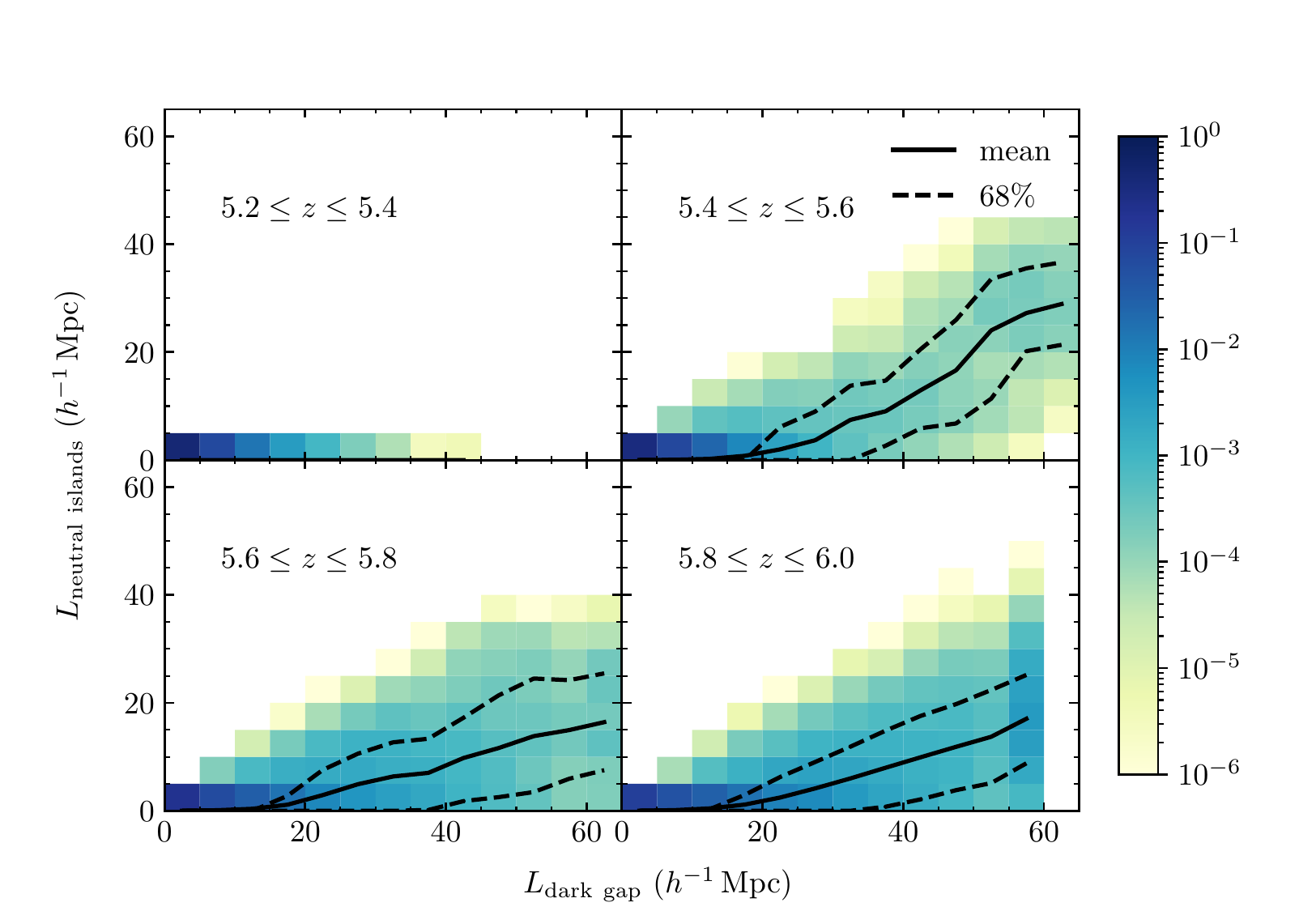}
    \caption{Correlation between neutral islands coverage and dark gap length in the {\tt K20-low-$\tau_{\rm CMB}$} model based on 10000 realizations. The histogram is calculated on $5\times5~(\cmpch)^2$ bins and color indicates the normalized probability, and all dark gaps are included regardless of whether they contain neutral gas. Solid and dashed lines show the mean and 68\% interval of the neutral islands coverage.}
    \label{fig:neutral_gap_L}
\end{figure*}

We further investigate the correlation between neutral islands coverage and dark gap length in the {\tt K20-low-$\tau_{\rm CMB}$} model at different redshift, as shown in Figure \ref{fig:neutral_gap_L}. The histogram is calculated based on 10000 realizations, and we include all dark gaps regardless of whether they contain neutral pixels.
The neutral islands coverage shown here is the sum of the  line-of-sight length of neutral pixels inside a dark gap. The mean neutral islands coverage is proportional to the dark gap length, meaning that long dark gaps may contain more neutral gas.  Nevertheless, the neutral islands coverage is, on average, significantly less than the dark gap length. This suggests that 
UVB fluctuations also play a significant role in producing the dark gaps in the late reionization models.

\section{Discussion \label{sec:discussion}}

\subsection{Model comparisons \label{sec:model_comparison}}

\begin{figure*}[ht]
    \centering
    \includegraphics[width=5in]{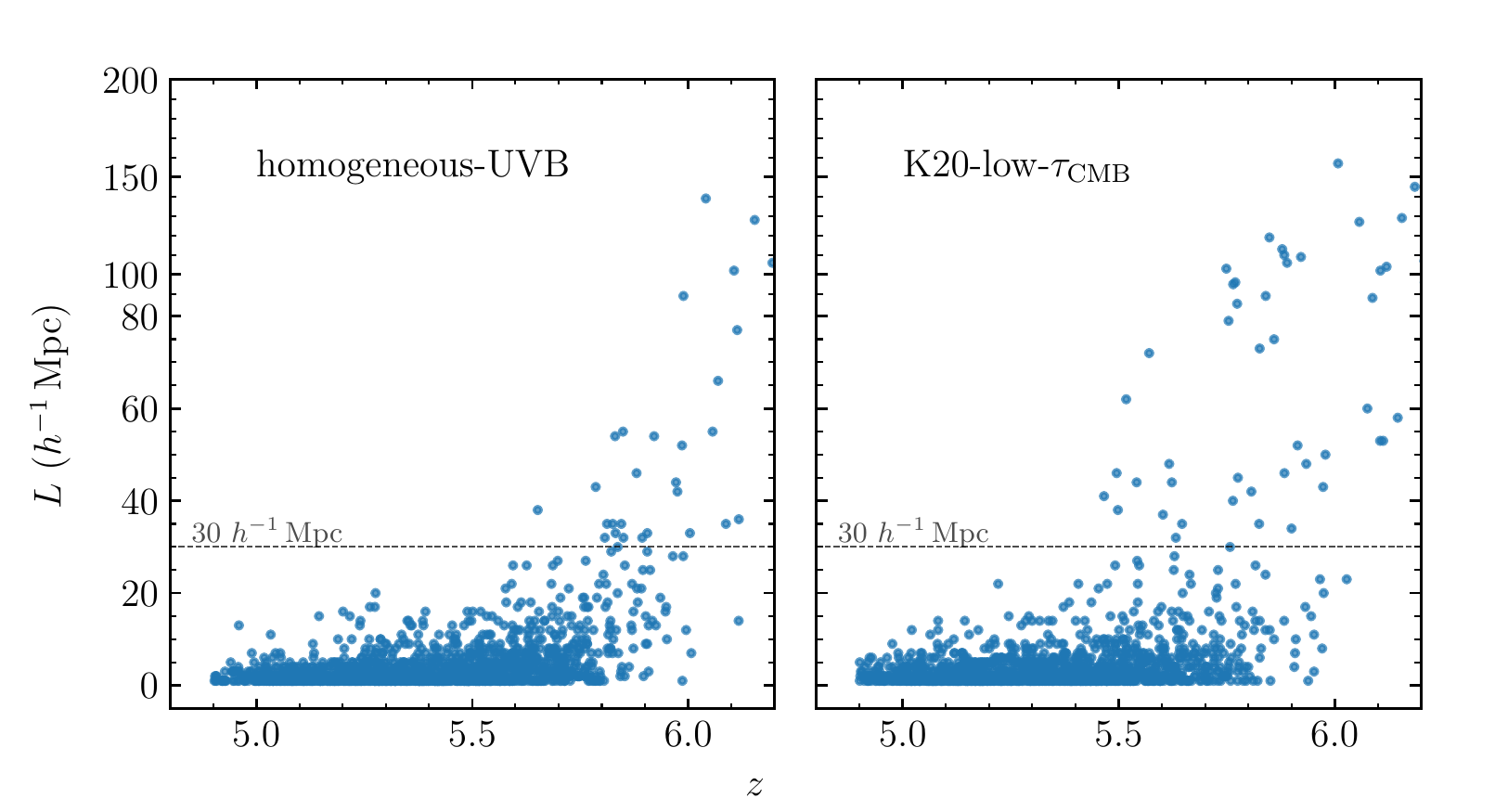}
    \caption{Gap length versus central redshift for dark gaps detected in mock spectra. For both models, the results shown here are based on one randomly selected set of mock spectra that matches our QSO sample in redshift and S/N ratio.}
    \label{fig:scatter-mock}
\end{figure*}

\begin{figure*}
    \centering
    \includegraphics[width=5.5in]{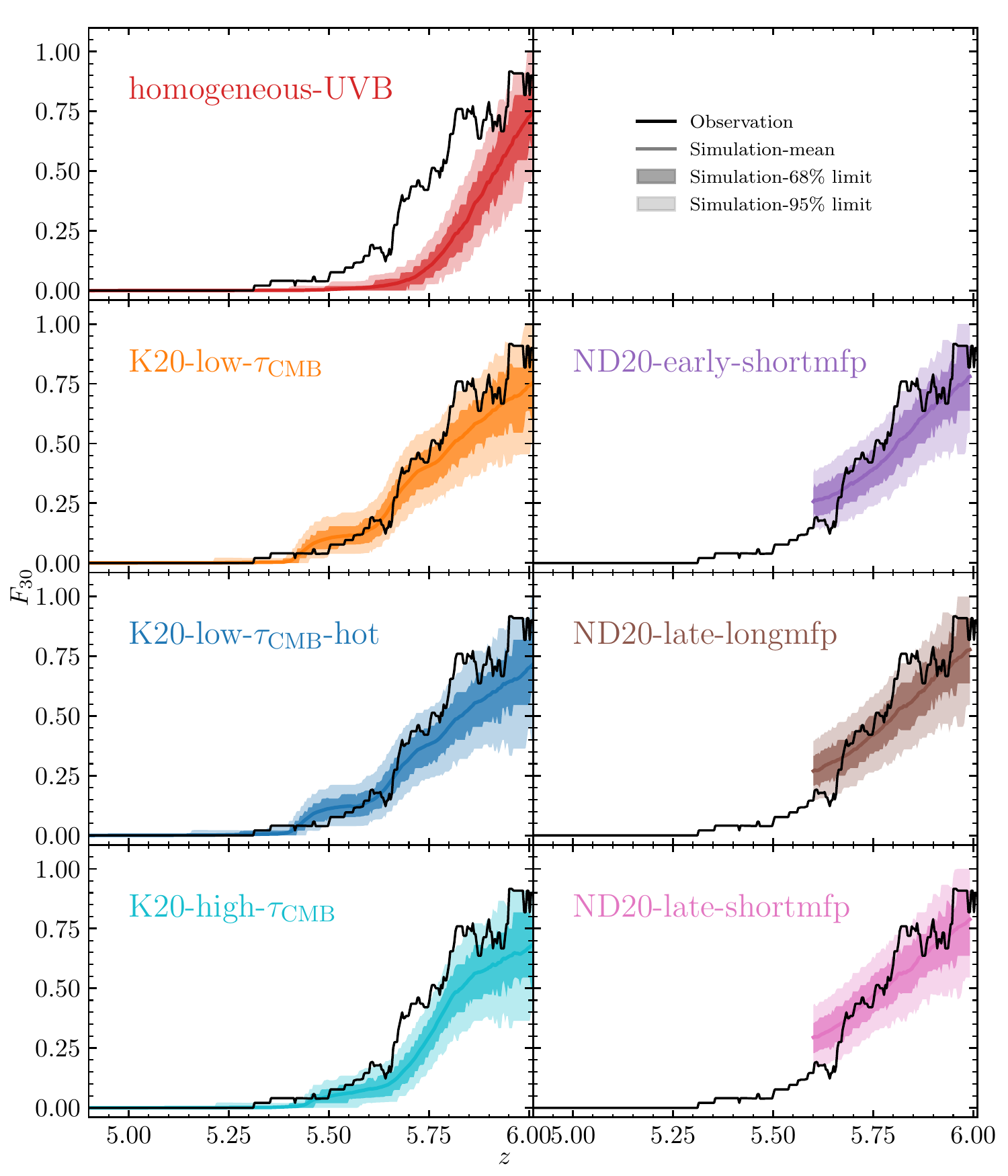}
    \caption{The fraction of sightlines located in dark gaps with
        $L \ge 30 \cmpch$ as a function of redshift.
        Both the observations (solid black line) and simulations (colored solid line and shaded regions) include gaps that are truncated at the red end of the \lya\ forest.
        Note that we discard dark gaps that are entirely in the 7 proper-Mpc proximity zone and/or in the buffer zone that covers $30\cmpch$ blueward the proximity zone cut.
        Dark and light shaded regions show the
        68\% and 95\% intervals, respectively, spanned by the models.}
    \label{fig:frac}
\end{figure*}

\begin{figure*}
    \centering
    \includegraphics[width=6.5in]{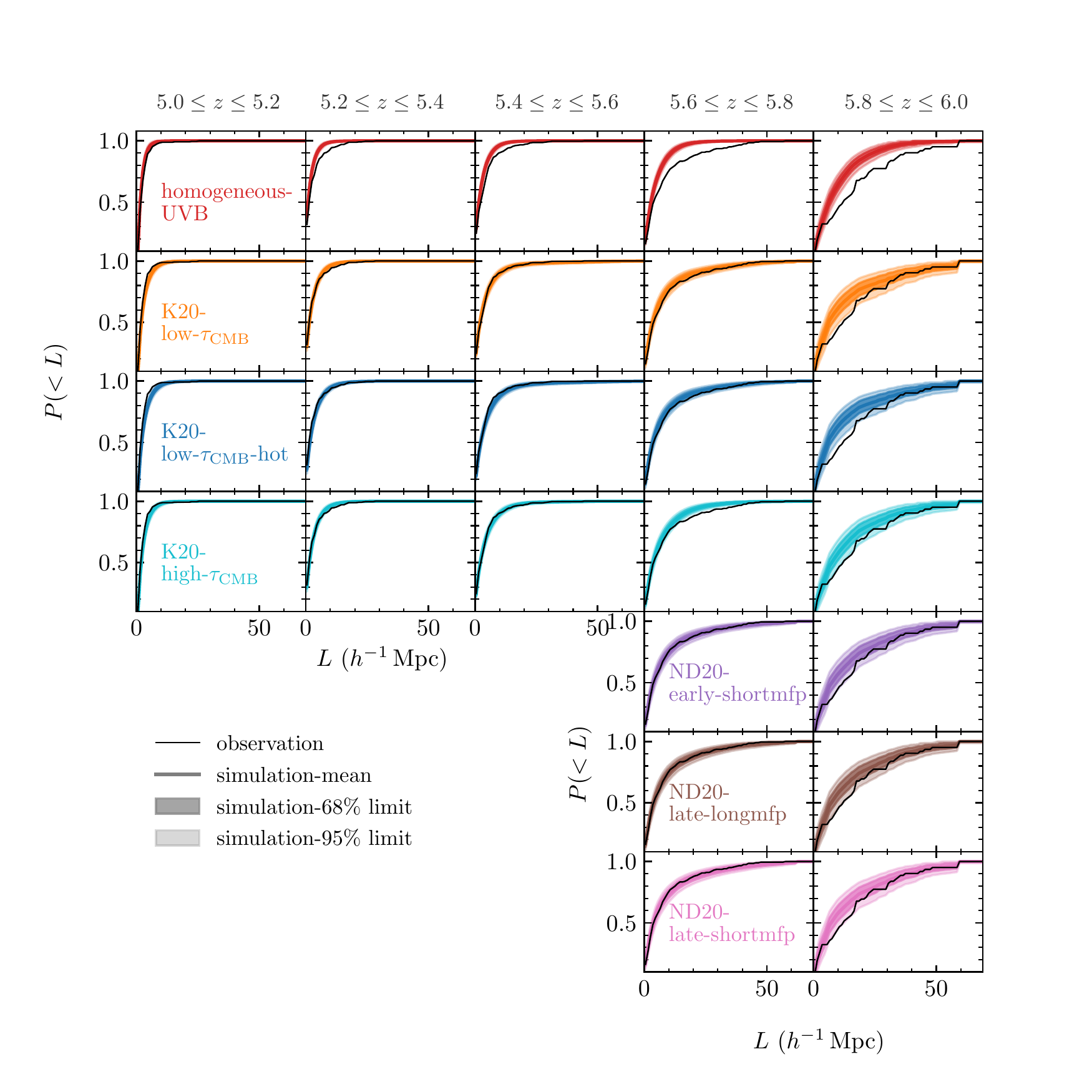}
    \caption{Cumulative distributions of dark gap length.
    In each redshift bin, the black line shows $P(<L)$ of the observed dark gaps. Dark gaps extending beyond the boundaries of the $\Delta z = 0.2$ window are truncated at the edge.
    The colored lines and dark/light shaded regions represent the mean and 68\%/95\% limits of $P(<L)$ in mock samples drawn from the models. 
        In this figure, we use a high-mass-resolution run from the Sherwood Simulation Suite \citep{bolton_sherwood_2017} for the {\tt homogeneous-UVB} model instead of the fiducial configuration. 
        See text for details.
    \label{fig:CDF}}
\end{figure*}

\begin{figure*}
    \centering
    \includegraphics[width=6.5in]{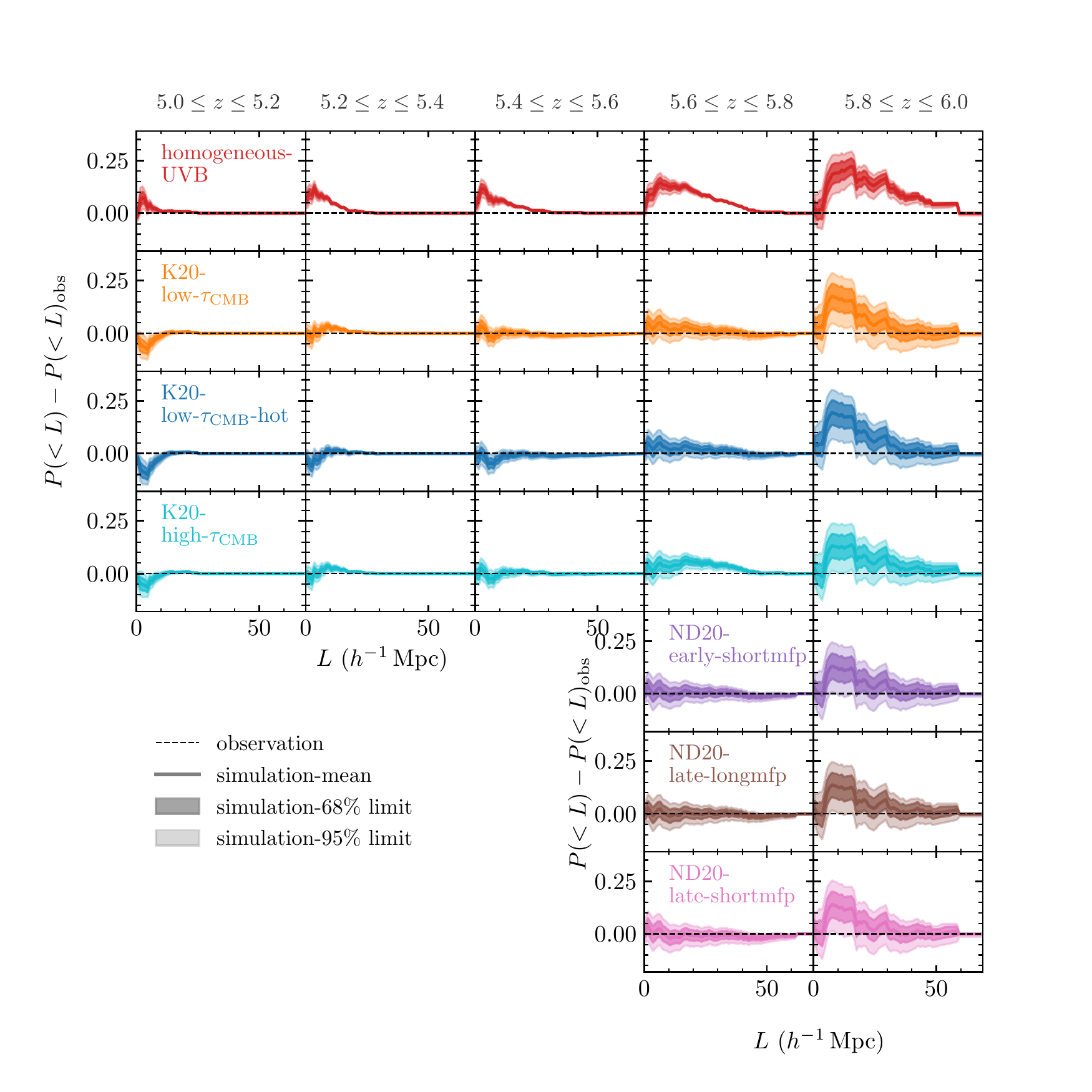}
    \caption{Similar to Figure \ref{fig:CDF}, but showing the differences on cumulative distributions of dark gap length between the models and the observations. 
    \label{fig:dCDF}}
\end{figure*}

We now compare our results to predictions from the simulations described in Section \ref{sec:simulations}. Figure \ref{fig:scatter-mock} plots the dark gap length versus central redshift 
for representative mock samples drawn from the {\tt homogeneous-UVB} model and the {\tt K20-low-$\tau_{\rm CMB}$} model. Qualitatively, as redshift increases, the {\tt homogeneous-UVB} model predicts a milder increase in long dark gaps than is seen in either the {\tt K20-low-$\tau_{\rm CMB}$} model or the observations (Figure \ref{fig:scatter}). To quantify the differences, we compute the relevant statistics by drawing mock samples from the simulations that match our observed QSO spectra in redshift and S/N ratio. 
We then compute the
dark gap statistics described in Section \ref{sec:dark_gap_stat}. We repeat this process 10000 times for each model and compute the mean, 68\% and 95\% limits on the expected scatter for the present sample size.  Figure \ref{fig:frac} compares $F_{30}$ predicted by models to that calculated from data. 
The jagged edges of the simulation confidence intervals are caused by the combined effects of step changes in the number of sightlines with redshift and the quantization of $F_{30}$ for a finite sample size.

The top left panel shows that the {\tt homogeneous-UVB} model is highly inconsistent with the observations over $5.3\lesssim z \lesssim 5.9$. At $z\sim 5.8$, the {\tt homogeneous-UVB} model under-predicts $F_{30}$ by a factor of three. At $z\sim 5.4$ and over $5.5 \lesssim z \lesssim 5.8$, this model is rejected by the data with $>99.9\%$ confidence. 

On the other hand, the {\tt K20-low-$\tau_{\rm CMB}$} and {\tt K20-low-$\tau_{\rm CMB}$-hot} models, wherein reionization ends at $z \simeq 5.3$, produce $F_{30}$ results that are generally consistent with the observations over $5<z<6$. One exception is that these models under-predict the small number of long dark gaps observed at $z\sim5.4$.  The {\tt K20-high-$\tau_{\rm CMB}$} model is consistent with the observations at $z \ge 5.75$ but under-predicts $F_{30}$ at lower redshifts.  This is a natural consequence of the earlier reionization in this model, which leads to a lower neutral hydrogen fraction and smaller UVB fluctuations at these redshifts. 

As shown in the right panels, $F_{30}$ values from the \citet{nasir_observing_2020} models are consistent with the observations within their 95\% limits over the available redshift range. Among the {\tt ND20} models, {\tt ND20-early-shortmfp} gives lower $F_{30}$ values compared to {\tt ND20-late}, but the difference is within the 68\% range for the present sample size.

We compare the cumulative distributions of dark gap length in Figure \ref{fig:CDF}, and give the differences between the observation and the model predictions in Figure \ref{fig:dCDF}.  In order to facilitate a direct comparison between the observations and simulations, we divide the data into redshift bins of $\Delta z = 0.2$. Here, dark gaps extending beyond the boundaries of a redshift bin are truncated at the edge when calculating $P(<L)$ for both the observation and models. Similar to our approach in Section \ref{sec:NL_PDF}, we do this to avoid artifacts from the finite length of the mock spectra. 

We present numerical convergence tests for the {\tt homogeneous-UVB} model in Appendix \ref{app: convergence}.  We find that the results for both $F_{30}$ and $P(<L)$ are relatively insensitive to box size, but that the number of small gaps increases with increasing mass resolution.
The impact of mass resolution is more significant for $P(<L)$ at smaller gap lengths than for $F_{30}$. For $P(<L)$ measured from the {\tt homogeneous-UVB} model, therefore, we display predictions based on a higher-resolution run with $2\times2048^3$ particles and a box size of $L=40\cmpch$ (hereafter {\tt 40\_2048}) instead of the fiducial configuration of $2\times2048^3$ particles and box size of $L=160\cmpch$ (hereafter {\tt 160\_2048}). Because \citet{keating_constraining_2020} use post-processed radiative transfer simulations, and \citet{nasir_observing_2020} simulations are based on an Eulerian code instead of a SPH code, mass resolution effects may be significantly different for these models than for the {\tt homogeneous-UVB} model.
We therefore present results as they are, although mass resolution corrections may be needed.

Over $z=5.6$--6.0, the {\tt homogeneous-UVB} model predicts significantly fewer long gaps than are observed in the data. The discrepancies between the data and the {\tt homogeneous-UVB} model persist down to the $z=5.2$--5.4 bin.

In contrast, the late reionization models, {\tt K20-low-$\tau_{\rm CMB}$}, {\tt K20-low-$\tau_{\rm CMB}$-hot}, and {\tt K20-high-$\tau_{\rm CMB}$}, predict $P(<L)$ values that are generally consistent with the data. Nevertheless, over $z=5.7$--5.9, we note that these models, especially the {\tt K20-high-$\tau_{\rm CMB}$} model, systematically yield higher $P(<L)$, i.e. fewer long gaps, than the observed for some $L$, though the discrepancies are less conspicuous compared to those for the {\tt homogeneous-UVB}. At lower redshifts, there are minor differences between the {\tt K20} models and the observation. The {\tt ND20-early-shortmfp} and {\tt ND20-late} models are generally consistent with the observation in the redshift range ($5.6\leq z \leq 6.0 $) currently probed by the simulations. 

\subsection{Implications for reionization}
Combining the results for $F_{30}$ and $P(<L)$, it is evident that a fully ionized IGM with a homogeneous UV background is disfavored by the observations down to $z\sim5.3$. 
This result is consistent with the large-scale inhomogeneities in IGM \lya\ opacity seen in recent $\teff$ measurements \citep[][]{becker_evidence_2015, bosman_new_2018,eilers_opacity_2018,yang_measurements_2020-1,bosman_hydrogen_2021}.

The late reionization models from \citet{keating_constraining_2020} and \citet{nasir_observing_2020} are generally consistent with dark gap 
statistics in the \lya\ forest. In these models, the residual neutral islands at $z < 6$ coupled with UVB fluctuations can naturally explain the appearance of long dark gaps in the \lya\ forest.
Among these models, the data tend to prefer those with later and more rapid reionization histories.  For example, the {\tt K20-low-$\tau_{\rm CMB}$} and {\tt K20-low-$\tau_{\rm CMB}$-hot} models, which have a reionization midpoint of $z_{50} = 6.7$, is somewhat more consistent (see curves and shades near $5.6\leq z \leq 5.8$ in Figures \ref{fig:frac}, \ref{fig:CDF}, and \ref{fig:dCDF}) with the dark gap statistics at $z < 6$ than the {\tt K20-high-$\tau_{\rm CMB}$} mode, for which $z_{50} = 8.4$.
A late and rapid reionization is also suggested by the recent mean free path measurement from \citet{becker_mean_2021} (see also \citealp{cain_short_2021,davies_predicament_2021}).

Alternatively, long dark gaps can arise from a fully reionized IGM provided that there are large UVB fluctuations.  The early reionization model from \citet{nasir_observing_2020}, which retains post-reionization fluctuations in the UV background and IGM temperature, is consistent with the data over at least $5.6 < z < 6.0$, where the available simulation outputs allow mock spectra to be compared to the data using the methods described above.
Extending these simulations down to lower redshifts would be helpful 
for testing the pure fluctuating UVB model further.

\section{Summary \label{sec:summary}}

In this paper, we present a search for dark gaps in the \lya\ forest over
$5 < z < 6$.  We use high-S/N spectra of 55 QSOs at $\zem>5.5$ taken with Keck ESI and VLT X-Shooter, including data from the new XQR-30 VLT Large Programme.  We focus on two statistics: the fraction of sightlines containing dark gaps of length $L \ge 30\cmpch$ as a function
of redshift, $F_{30}$, which we introduce here for the first time, and the dark gap length distribution, $P(<L)$. 
Our primary goal is to quantify the persistence of large \lya-opaque regions in the IGM below redshift six, and to evaluate the consistency between the observed dark gap statistics and predictions from various models.  We include a model with a fully ionized IGM and a uniform ionizing UV background, and others with large islands of neutral gas and/or UVB fluctuations.  Our main results can be summarized as follows:

\begin{itemize}

    \item We identify \add{50} long dark gaps ($L \ge 30\cmpch$) in the \lya\ forest from our sample. Two long dark gaps are found at $z<5.5$, with one extending down to $z\simeq 5.3$. We also report new ultra-long dark gaps ($L>80\cmpch$) below $z=6$, similar to the one previously reported towards ULAS J0148+0600 by \citet{becker_evidence_2015}. The presence of long dark gaps at these redshifts demonstrates that large regions of the IGM remain opaque to \lya\ down to $z\simeq 5.3$.  
        \item In terms of both $F_{30}$ and $P(<L)$, a fully ionized IGM with a homogeneous UVB is disfavored by the data down to $z\simeq5.3$.
    
    \item Models wherein reionization ends significantly below redshift six \citep[][]{keating_constraining_2020,nasir_observing_2020} are broadly consistent with the data. Among these, the data favor models with a reionization midpoint near $z \sim 7$ and an end at $z \simeq 5.3$ or later.
        In these models, dark gaps arise from a combination of neutral patches in the IGM and regions of low ionizing UV background, which are often adjacent to one another.

    \item We also find consistency with a model wherein reionization ends by $z = 6$ but the IGM retains large fluctuations in the UV background \citep[][]{nasir_observing_2020},
    at least over $5.6 < z < 6.0$.
        
\end{itemize}

Overall, the evolution of dark gaps observed at $z<6$ suggests that signatures of reionization remain present in the IGM until at least $z\simeq 5.3$ in the form of neutral hydrogen islands and/or fluctuations in the ionizing UV background.
We note that this 
work focuses on dark gaps in the \lya\ forest. Given its lower optical depth, however, \lyb\ may also be a useful tool. For example, islands of neutral gas may tend to produce more long \lyb\ troughs than are created by fluctuations in the UV background \citep[e.g.,][]{nasir_observing_2020}.
These and other statistics should provide further details on how the IGM evolves near the end of reionization.

\begin{acknowledgments}

We thank Elisa Boera and Fahad Nasir for their help and useful discussion. \add{We also thank the anonymous reviewer for their careful reading of the paper and thoughtful comments.}

YZ, GDB, and HMC were supported by the National Science Foundation through grants AST-1615814 and AST-1751404. HMC was also supported by the National Science Foundation Graduate Research Fellowship Program under Grant No. DGE-1326120.
SEIB acknowledges funding from the European Research Council (ERC) under the European Union's Horizon 2020 research and innovation programme (grant agreement No. 740246 ``Cosmic Gas'').
LCK was supported by the European Union's Horizon 2020 research and innovation programme under the Marie Skłodowska-Curie grant agreement No. 885990.
FB acknowledges support from the Australian Research Council through Discovery Projects (award DP190100252) and Chinese Academy of Sciences (CAS) through a China-Chile Joint Research Fund (CCJRF1809) administered by the CAS South America Center for Astronomy (CASSACA).
ACE acknowledges support by NASA through the NASA Hubble Fellowship grant $\#$HF2-51434 awarded by the Space Telescope Science Institute, which is operated by the Association of Universities for Research in Astronomy, Inc., for NASA, under contract NAS5-26555. 
XF and JY acknowledge support from the NSF grants AST 15-15115 and AST 19-08284. 
MGH acknowledges support from the UKRI STFC (grant numbers ST/N000927/1 and ST/S000623/1).  
GK’s research is partly supported by the Max Planck Society via a partner group grant.
AP acknowledges support from the ERC Advanced Grant INTERSTELLAR H2020/740120.
Parts of this work was supported by the Australian Research Council Centre of Excellence for All Sky Astrophysics in 3 Dimensions (ASTRO 3D), through project $\#$CE170100013.
FW thanks the support provided by NASA through the NASA Hubble Fellowship grant \#HST-HF2-51448.001-A awarded by the Space Telescope Science Institute, which is operated by the Association of Universities for Research in Astronomy, Incorporated, under NASA contract NAS5-26555.

Based on observations collected at the European Southern Observatory under ESO programmes 060.A-9024(A), 084.A-0360(A), 084.A-0390(A), 084.A-0550(A), 085.A-0299(A), 086.A-0162(A),  
086.A-0574(A), 087.A-0607(A), 088.A-0897(A), 091.C-0934(B), 096.A-0095(A), 096.A-0418(A), 
097.B-1070(A), 098.A-0111(A), 098.B-0537(A), 0100.A-0243(A), 0100.A-0625(A), 0101.B-0272(A), 0102.A-0154(A), 0102.A-0478(A), 1103.A-0817(A), and 1103.A-0817(B).

Some of the data presented herein were obtained at the W. M. Keck Observatory, which is operated as a scientific partnership among the California Institute of Technology, the University of California and the National Aeronautics and Space Administration. The Observatory was made possible by the generous financial support of the W. M. Keck Foundation.
The authors wish to recognize and acknowledge the very significant cultural role and reverence that the summit of Maunakea has always had within the indigenous Hawaiian community. We are most fortunate to have the opportunity to conduct observations from this mountain. Finally, this research has made use of the Keck Observatory Archive (KOA), which is operated by the W.M. Keck Observatory and the NASA Exoplanet Science Institute (NExScI), under contract with the National Aeronautics and Space Administration.

This work was performed using the Cambridge Service for Data Driven Discovery (CSD3), part of which is operated by the University of Cambridge Research Computing on behalf of the STFC DiRAC HPC Facility (www.dirac.ac.uk). The DiRAC component of CSD3 was funded by BEIS capital funding via STFC capital grants ST/P002307/1 and ST/R002452/1 and STFC operations grant ST/R00689X/1. This work further used the DiRAC@Durham facility managed by the Institute for Computational Cosmology on behalf of the STFC DiRAC HPC Facility. The equipment was funded by BEIS capital funding via STFC capital grants ST/P002293/1 and ST/R002371/1, Durham University and STFC operations grant ST/R000832/1. DiRAC is part of the National e-Infrastructure. 

\end{acknowledgments}

\vspace{5mm}
\facilities{Keck:II (ESI), VLT:Kueyen (X-Shooter)}

\software{
    {\tt Astropy} \citep{astropy_collaboration_astropy_2013,astropy_collaboration_astropy_2018},
    {\tt Matplotlib} \citep{hunter_matplotlib_2007},
    {\tt NumPy} \citep{harris_array_2020},
    {\tt SpectRes} \citep{carnall_spectres_2017}
}

\clearpage
\appendix
\restartappendixnumbering

\section{Numerical Convergence}\label{app: convergence}

\begin{figure*}[ht]
    \centering
    \includegraphics[width=7in]{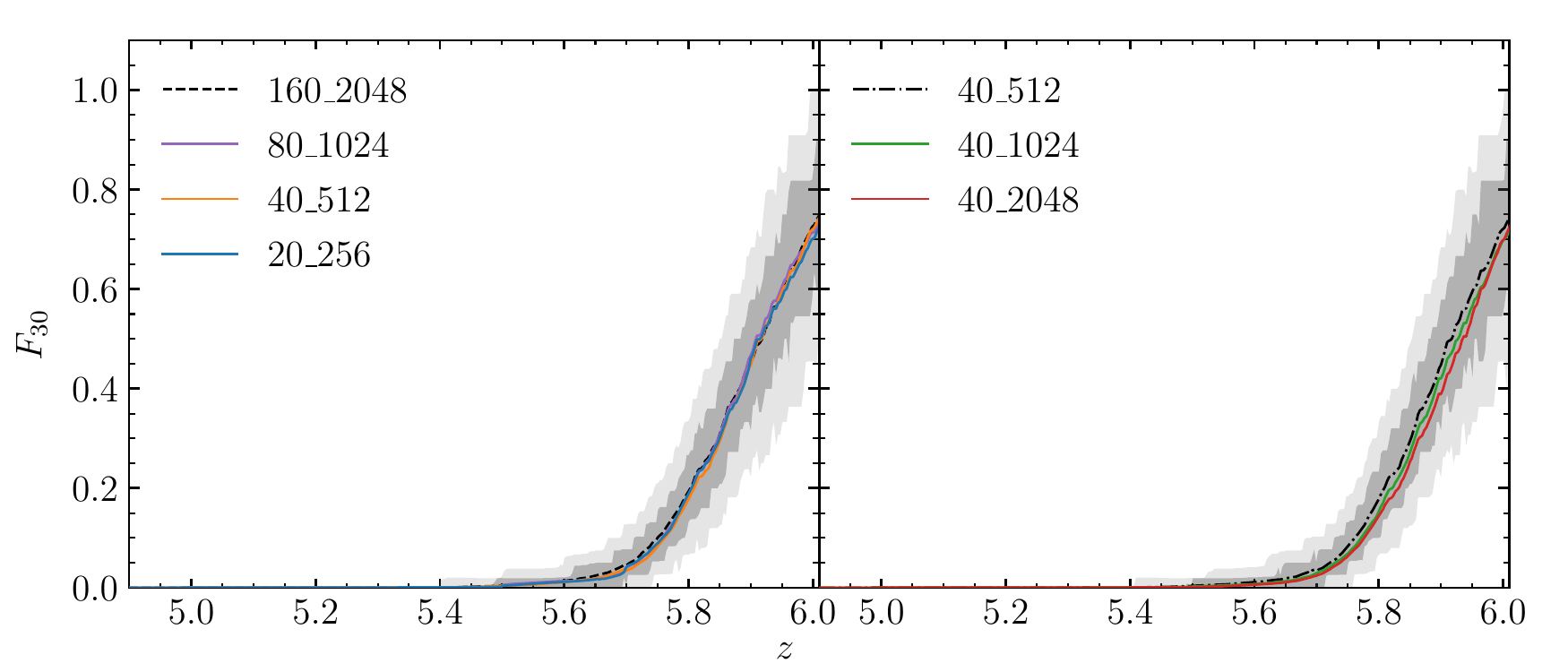}
    \caption{Fractions of QSO spectra exhibiting long ($L \geq 30\cmpch$) dark gaps as a function of redshift with different simulation configurations.
        The left panel compares results for varying box sizes but fixed
        mass resolution. The dark gray and light gray shaded regions are the 68\% and 95\% limits on the expected scatter for the present sample size from the {\tt 160\_2048} simulation, which are the same as the shaded regions in Figure \ref{fig:frac}.
        The right panel compares results for varying mass resolutions but fixed
        box size. The dash dotted line is the mean, and shaded regions are 68\% and 95\% limits of the prediction from the {\tt 40\_512} simulation.
        \label{fig:conv_f30}  
                }
\end{figure*}

\begin{figure*}[ht]
    \centering
    \includegraphics[width=7.1in]{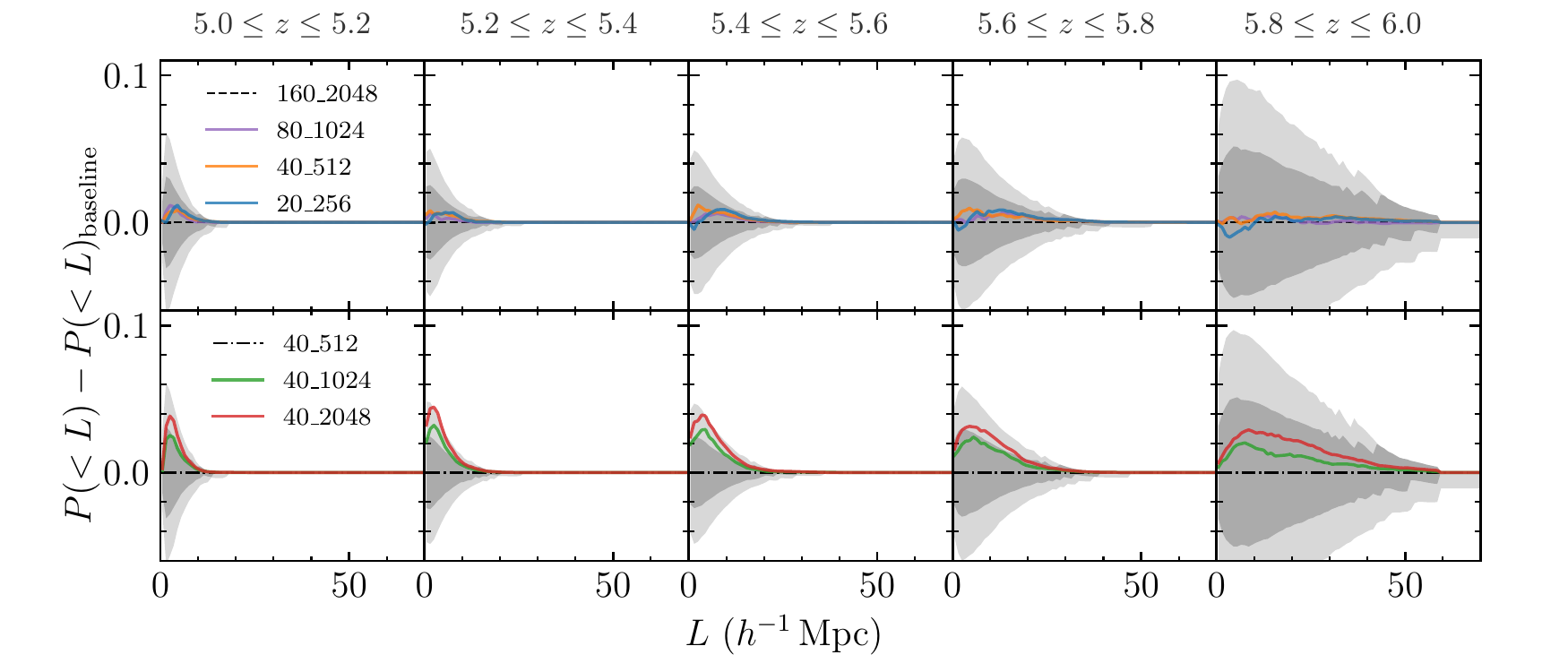}
    \caption{Difference on dark gap length distributions for different box sizes and mass resolutions compared to the baseline configurations. 
    The top and bottom rows compare results for varying box sizes and varying mass resolutions, respectively. The dark gray and light gray shaded regions are the 68\% and 95\% limits on the expected scatter for the present sample size from the {\tt 160\_2048} simulation (Top panel) and from the {\tt 40\_512} simulation (Bottom panel). 
        \label{fig:conv_CDF}
    }
\end{figure*}

Here we test the convergence of our results for the {\tt homogeneous-UVB} model with different box sizes and mass resolutions. We denote different simulation configurations as {\tt X\_Y}, where X is the box size in comoving ${\rm Mpc}~h^{-1}$ and ${\rm Y} =  [({\rm number~of~dark~matter + baryon~particles})/2]^{1/3}$. The fiducial configuration used in Section \ref{sec:dark_gap_stat} is {\tt 160\_2048}.

\add{To calculate $F_{30}$ with smaller boxes, we firstly stitch the short skewers to form  
160 Mpc/h skewers, and then create mock spectra following the method described in Section \ref{sec:mock}.}
In Figure \ref{fig:conv_f30}, we compare $F_{30}$ for mock data generated from different simulation configurations. We find little dependence on box size.  $F_{30}$ decreases slightly with increasing mass resolution, though the differences are within the expected 68\% scatter for the present sample size. We compare $P(<L)$ with different configurations to the baseline configurations by calculating $P(<L) - P(<L)_{\rm baseline}$ in Figure \ref{fig:conv_CDF}. For a fixed mass resolution and varying box size, we compare simulations to our fiducial {\tt 160\_2048} simulation.  For a fixed box size but varying mass resolution, we compare simulations to the {\tt 40\_512} configuration, which has the same mass resolution of {\tt 160\_2048}. Similar to $F_{30}$, the variations in $P(<L)$ with box size are relatively minor.  The impact of mass resolution is more significant, especially for smaller $L$.

Figures \ref{fig:conv_f30} and \ref{fig:conv_CDF} suggest that mass resolution has a larger impact than box size on our statistics, in the sense that simulations with lower mass resolutions tend to produce more long gaps and fewer short gaps. This is because weak, narrow \lya\ transmission peaks tend to be suppressed at lower resolution.
This effect may need to be considered for larger samples.  We emphasize that the {\tt homogeneous-UVB} models that we are using for these convergence tests contain significantly fewer long gaps than the late reionization and fluctuating UVB models.  It is therefore unclear how well the trends we see for large $L$ would apply to these models, although it is likely that the effects of mass resolution we see at smaller gap lengths would generally be present in SPH simulations. 

\section{Simulation Predictions without Masking} \label{app:no_masking}
\restartappendixnumbering

\begin{figure*}[ht]
    \centering
    \includegraphics[width=5.5in]{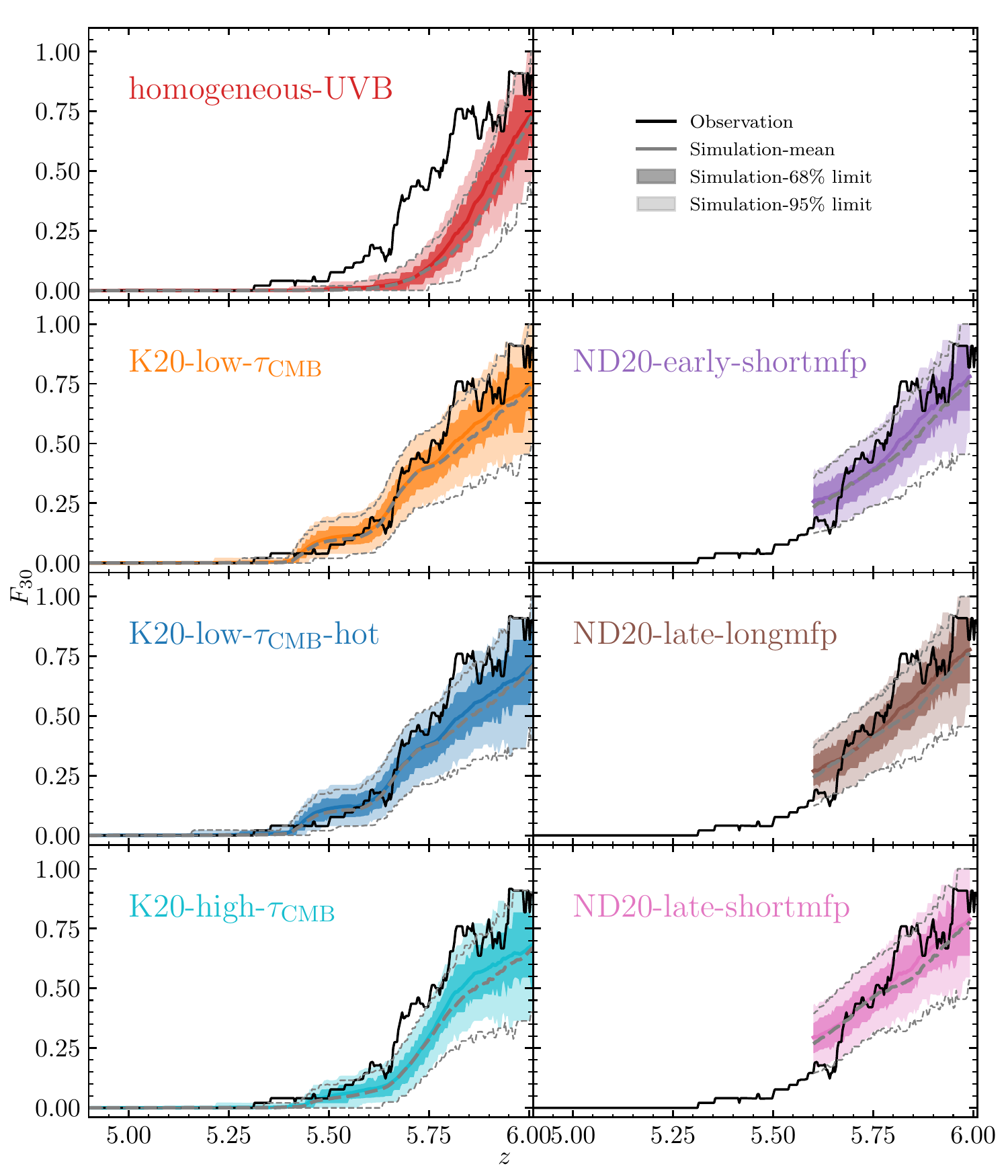}
    \caption{The fraction of sightlines located in dark gaps with
        $L \ge 30 \cmpch$ as a function of redshift. The dashed thick (thin) lines show the mean values (95\% range) of simulation predictions without masking regions in the mock spectra that coincide with peaks in the flux error array.
        Other lines and shaded regions are as described in Figure \ref{fig:frac}. }
    \label{fig:frac_no_masking}
\end{figure*}

For consistency with the observations,
we mask out small wavelength regions in the mock spectra that coincide with peaks in the observed flux error arrays, as described in Section \ref{sec:methods}.
Figure \ref{fig:frac_no_masking} shows how the simulation results change without masking. The model predictions for $F_{30}$ decrease because the masks sometimes fall on transmission peaks.  The overall impact is minor; however, we emphasize that the observations should be compared to the simulation results with masking included.

\section{$F_{30}$ based on the power law continuum fitting \label{app:powerlaw}}
\begin{figure}[ht]
    \begin{center}
        \includegraphics[width=3.355in]{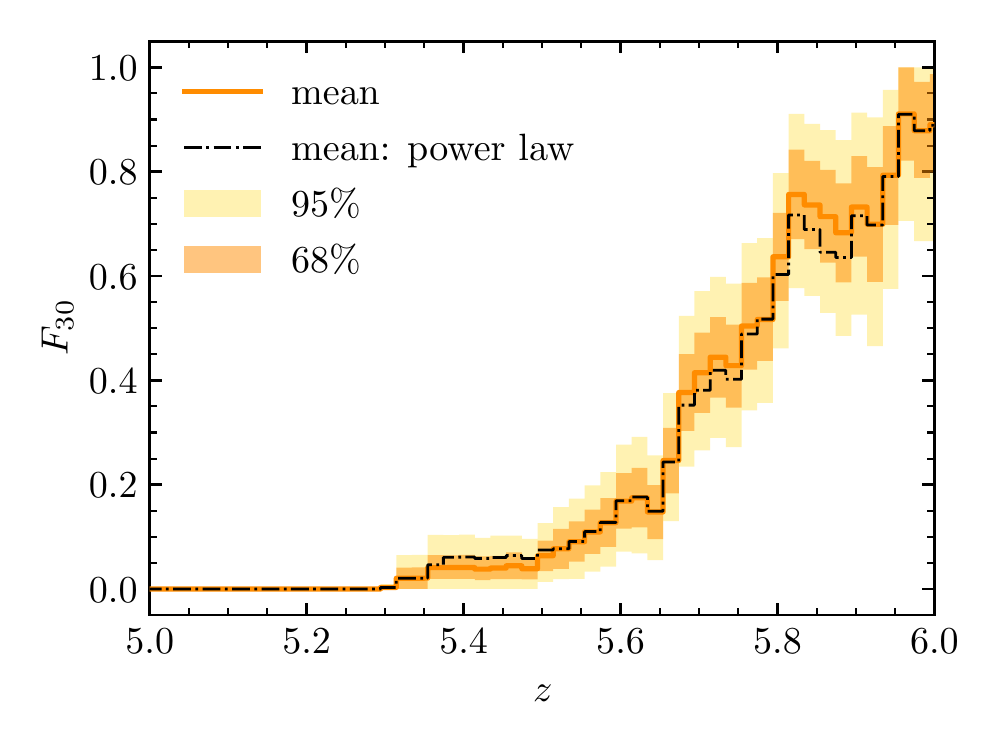}
        \caption{Measured fraction of QSO spectra exhibiting long ($L \ge 30 \cmpch$) dark gaps as a function of redshift. The dot dashed line shows the result based on the power law continuum fitting. Other lines and regions are as described in Figure \ref{fig:F30-obs}.}
        \label{fig:F30-obs-powerlaw}
    \end{center}
\end{figure}

\restartappendixnumbering
For reference, in this section, we calculate $F_{30}$ based on the power law continuum fitting. The power-law continua are in the form of $a\lambda^{-b}$, with $a$ and $b$ being free parameters. We generally  estimate the power-law continua over $\sim$1285-1350 \AA\ in the rest frame, which is relatively free of emission lines, and we extend the fitting range out to $\sim 2000$ \AA\ when possible. Figure \ref{fig:F30-obs-powerlaw} compares the results. The consistency (within $1\sigma$) between $F_{30}$ based on the PCA continuum and power law continuum suggests that our results are insensitive to continuum fitting methods.

\add{
\section{Effect metal absorbers on $F_{30}$ \label{app:nodla}}
\restartappendixnumbering
The strong \hi\ absorption typically associated with metal systems may potentially impact the observed $F_{30}$ by connecting otherwise shorter gaps. We test whether this effect could be significant by dividing dark gaps at the redshifts of DLAs and other metal systems.  We also exclude a $3000~\rm km\,s^{-1}$ region surrounding the redshift of the metal absorber in order to allow for extended DLA absorption and/or strong absorbers clustered around the  metal system.  
As shown in Figure \ref{fig:F30nodla}, the impact on $F_{30}$ is relativel minor, with a maximum decrease of $\sim$0.1 at $z \sim 5.8$.  We caution that list of metal absorbers used here may be incomplete; however, we have verified that the three long dark gaps at $z \lesssim 5.5$ in particular do not contain metals to within the sensitivity of our data.  In summary, we find that the impact of metal systems on $F_{30}$ in this regard is minor, and that the {\tt homogenous-UVB} model is strongly ruled out regardless of how these systems are treated.
\begin{figure}[ht]
    \begin{center}
        \includegraphics[width=5.5in]{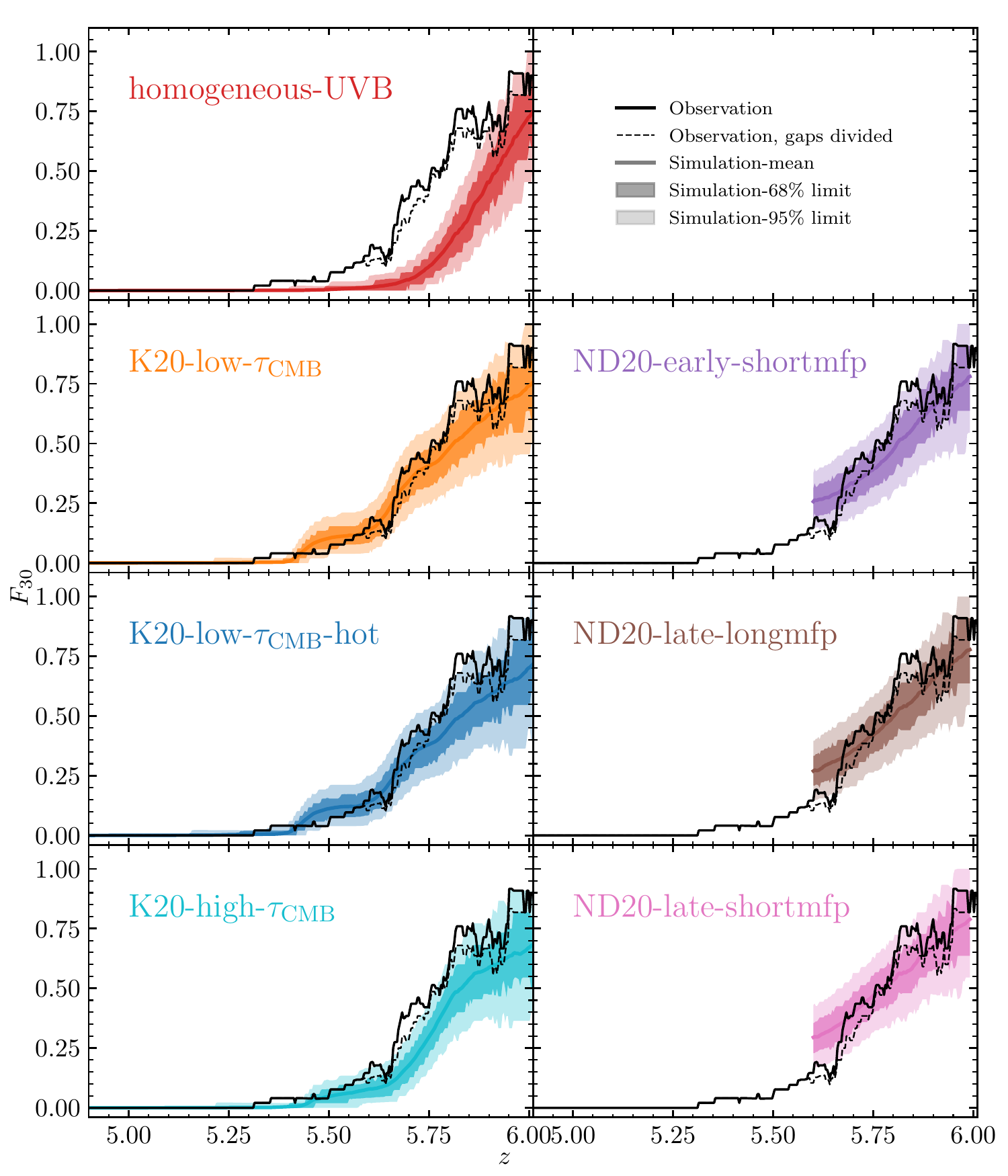}
        \caption{The fraction of sightlines located in dark gaps with
        $L \ge 30 \cmpch$ as a function of redshift. The dashed lines show $F_{30}$ from observations with dark gaps divided at the redshifts of DLAs or metal systems with a $3000~\rm km\,s^{-1}$ interval.
        Other lines and shaded regions are as described in Figure \ref{fig:frac}. }
    \label{fig:F30nodla}
    \end{center}
\end{figure}
}


\clearpage


\begin{thebibliography}{85}
\expandafter\ifx\csname natexlab\endcsname\relax\def\natexlab#1{#1}\fi

\bibitem[{{Astropy Collaboration} {et~al.}(2018){Astropy Collaboration},
  {Price-Whelan}, Sip{\H o}cz, G{\"u}nther, Lim, Crawford, Conseil, Shupe,
  Craig, Dencheva, Ginsburg, VanderPlas, Bradley, {P{\'e}rez-Su{\'a}rez}, {de
  Val-Borro}, Aldcroft, Cruz, Robitaille, Tollerud, Ardelean, Babej, Bach,
  Bachetti, Bakanov, Bamford, Barentsen, Barmby, Baumbach, Berry, Biscani,
  Boquien, Bostroem, Bouma, Brammer, Bray, Breytenbach, Buddelmeijer, Burke,
  Calderone, Cano~Rodr{\'i}guez, Cara, Cardoso, Cheedella, Copin, Corrales,
  Crichton, D'Avella, Deil, Depagne, Dietrich, Donath, Droettboom, Earl, Erben,
  Fabbro, Ferreira, Finethy, Fox, Garrison, Gibbons, Goldstein, Gommers, Greco,
  Greenfield, Groener, Grollier, Hagen, Hirst, Homeier, Horton, Hosseinzadeh,
  Hu, Hunkeler, Ivezi{\'c}, Jain, Jenness, Kanarek, Kendrew, Kern, Kerzendorf,
  Khvalko, King, Kirkby, Kulkarni, Kumar, Lee, Lenz, Littlefair, Ma, Macleod,
  Mastropietro, McCully, Montagnac, Morris, Mueller, Mumford, Muna, Murphy,
  Nelson, Nguyen, Ninan, N{\"o}the, Ogaz, Oh, Parejko, Parley, Pascual, Patil,
  Patil, Plunkett, Prochaska, Rastogi, Reddy~Janga, Sabater, Sakurikar,
  Seifert, Sherbert, {Sherwood-Taylor}, Shih, Sick, Silbiger, Singanamalla,
  Singer, Sladen, Sooley, Sornarajah, Streicher, Teuben, Thomas, Tremblay,
  Turner, Terr{\'o}n, {van Kerkwijk}, {de la Vega}, Watkins, Weaver, Whitmore,
  Woillez, Zabalza, \& {Astropy
  Contributors}}]{astropy_collaboration_astropy_2018}
{Astropy Collaboration}, {Price-Whelan}, A.~M., Sip{\H o}cz, B.~M., {et~al.}
  2018, \href{http://dx.doi.org/10.3847/1538-3881/aabc4f}{\color{magenta}\aj},
  \href{http://adsabs.harvard.edu/abs/2018AJ....156..123A}{156, 123}

\bibitem[{{Astropy Collaboration} {et~al.}(2013){Astropy Collaboration},
  Robitaille, Tollerud, Greenfield, Droettboom, Bray, Aldcroft, Davis,
  Ginsburg, {Price-Whelan}, Kerzendorf, Conley, Crighton, Barbary, Muna,
  Ferguson, Grollier, Parikh, Nair, Unther, Deil, Woillez, Conseil, Kramer,
  Turner, Singer, Fox, Weaver, Zabalza, Edwards, Azalee~Bostroem, Burke, Casey,
  Crawford, Dencheva, Ely, Jenness, Labrie, Lim, Pierfederici, Pontzen, Ptak,
  Refsdal, Servillat, \& Streicher}]{astropy_collaboration_astropy_2013}
{Astropy Collaboration}, Robitaille, T.~P., Tollerud, E.~J., {et~al.} 2013,
  \href{http://dx.doi.org/10.1051/0004-6361/201322068}{\color{magenta}\aap},
  \href{http://adsabs.harvard.edu/abs/2013A%26A...558A..33A}{558, A33}

\bibitem[{Aubert \& Teyssier(2008)}]{aubert_radiative_2008}
Aubert, D. \& Teyssier, R. 2008,
  \href{http://dx.doi.org/10.1111/j.1365-2966.2008.13223.x}{\color{magenta}\mnras},
  \href{http://adsabs.harvard.edu/abs/2008MNRAS.387..295A}{387, 295}

\bibitem[{Aubert \& Teyssier(2010)}]{aubert_reionization_2010}
---. 2010,
  \href{http://dx.doi.org/10.1088/0004-637X/724/1/244}{\color{magenta}\apj},
  \href{http://adsabs.harvard.edu/abs/2010ApJ...724..244A}{724, 244}

\bibitem[{Ba{\~n}ados {et~al.}(2015)Ba{\~n}ados, Decarli, Walter, Venemans,
  Farina, \& Fan}]{banados_bright_2015}
Ba{\~n}ados, E., Decarli, R., Walter, F., {et~al.} 2015,
  \href{http://dx.doi.org/10.1088/2041-8205/805/1/L8}{\color{magenta}\apjl},
  \href{http://adsabs.harvard.edu/abs/2015ApJ...805L...8B}{805, L8}

\bibitem[{Ba{\~n}ados {et~al.}(2018)Ba{\~n}ados, Venemans, Mazzucchelli,
  Farina, Walter, Wang, Decarli, Stern, Fan, Davies, Hennawi, Simcoe, Turner,
  Rix, Yang, Kelson, Rudie, \& Winters}]{banados_800-million-solar-mass_2018}
Ba{\~n}ados, E., Venemans, B.~P., Mazzucchelli, C., {et~al.} 2018,
  \href{http://dx.doi.org/10.1038/nature25180}{\color{magenta}\nat},
  \href{http://adsabs.harvard.edu/abs/2018Natur.553..473B}{553, 473}

\bibitem[{Barnett {et~al.}(2017)Barnett, Warren, Becker, Mortlock, Hewett,
  McMahon, Simpson, \& Venemans}]{barnett_observations_2017}
Barnett, R., Warren, S.~J., Becker, G.~D., {et~al.} 2017,
  \href{http://dx.doi.org/10.1051/0004-6361/201630258}{\color{magenta}\aap},
  \href{http://adsabs.harvard.edu/abs/2017A%26A...601A..16B}{601, A16}

\bibitem[{Becker {et~al.}(2015)Becker, Bolton, Madau, Pettini, {Ryan-Weber}, \&
  Venemans}]{becker_evidence_2015}
Becker, G.~D., Bolton, J.~S., Madau, P., {et~al.} 2015,
  \href{http://dx.doi.org/10.1093/mnras/stu2646}{\color{magenta}\mnras},
  \href{http://adsabs.harvard.edu/abs/2015MNRAS.447.3402B/abstract}{447, 3402}

\bibitem[{Becker {et~al.}(2021)Becker, D'Aloisio, Christenson, Zhu, Worseck, \&
  Bolton}]{becker_mean_2021}
Becker, G.~D., D'Aloisio, A., Christenson, H.~M., {et~al.}
  \href{http://adsabs.harvard.edu/abs/2021arXiv210316610B}{2021,
  arXiv:2103.16610}

\bibitem[{Becker {et~al.}(2018)Becker, Davies, Furlanetto, Malkan, Boera, \&
  Douglass}]{becker_evidence_2018}
Becker, G.~D., Davies, F.~B., Furlanetto, S.~R., {et~al.} 2018,
  \href{http://dx.doi.org/10.3847/1538-4357/aacc73}{\color{magenta}\apj},
  \href{http://adsabs.harvard.edu/abs/2018ApJ...863...92B}{863, 92}

\bibitem[{Becker {et~al.}(2019)Becker, Pettini, Rafelski, D'Odorico, Boera,
  Christensen, Cupani, Ellison, Farina, Fumagalli, L{\'o}pez, Neeleman,
  {Ryan-Weber}, \& Worseck}]{becker_evolution_2019}
Becker, G.~D., Pettini, M., Rafelski, M., {et~al.} 2019,
  \href{http://dx.doi.org/10.3847/1538-4357/ab3eb5}{\color{magenta}\apj},
  \href{http://adsabs.harvard.edu/abs/2019ApJ...883..163B}{883, 163}

\bibitem[{Boera {et~al.}(2019)Boera, Becker, Bolton, \&
  Nasir}]{boera_revealing_2019}
Boera, E., Becker, G.~D., Bolton, J.~S., \& Nasir, F. 2019,
  \href{http://dx.doi.org/10.3847/1538-4357/aafee4}{\color{magenta}\apj},
  \href{http://adsabs.harvard.edu/abs/2019ApJ...872..101B/abstract}{872, 101}

\bibitem[{Bolton {et~al.}(2017)Bolton, Puchwein, Sijacki, Haehnelt, Kim,
  Meiksin, Regan, \& Viel}]{bolton_sherwood_2017}
Bolton, J.~S., Puchwein, E., Sijacki, D., {et~al.} 2017,
  \href{http://dx.doi.org/10.1093/mnras/stw2397}{\color{magenta}\mnras},
  \href{http://adsabs.harvard.edu/abs/2017MNRAS.464..897B}{464, 897}

\bibitem[{Bosman {et~al.}(2021{\natexlab{a}})Bosman, Davies, Becker, Keating,
  Davies, Zhu, Eilers, D'Odorico, Bian, Bischetti, Cristiani, Fan, Farina,
  Haehnelt, Kulkarni, Mesinger, Meyer, Onoue, Pallottini, Qin, {Ryan-Weber},
  Schindler, Walter, Wang, \& Yang}]{bosman_hydrogen_2021}
Bosman, S. E.~I., Davies, F.~B., Becker, G.~D., {et~al.}
  \href{http://adsabs.harvard.edu/abs/2021arXiv210803699B}{2021{\natexlab{a}},
  arXiv:2108.03699}

\bibitem[{Bosman {et~al.}(2021{\natexlab{b}})Bosman, {\v D}urov{\v
  c}{\'i}kov{\'a}, Davies, \& Eilers}]{bosman_comparison_2021}
Bosman, S. E.~I., {\v D}urov{\v c}{\'i}kov{\'a}, D., Davies, F.~B., \& Eilers,
  A.~C. 2021{\natexlab{b}},
  \href{http://dx.doi.org/10.1093/mnras/stab572}{\color{magenta}\mnras},
  \href{http://adsabs.harvard.edu/abs/2021MNRAS.503.2077B}{503, 2077}

\bibitem[{Bosman {et~al.}(2018)Bosman, Fan, Jiang, Reed, Matsuoka, Becker, \&
  Haehnelt}]{bosman_new_2018}
Bosman, S. E.~I., Fan, X., Jiang, L., {et~al.} 2018,
  \href{http://dx.doi.org/10.1093/mnras/sty1344}{\color{magenta}\mnras},
  \href{http://adsabs.harvard.edu/abs/2018MNRAS.479.1055B}{479, 1055}

\bibitem[{Bromm \& Larson(2004)}]{bromm_first_2004}
Bromm, V. \& Larson, R.~B. 2004,
  \href{http://dx.doi.org/10.1146/annurev.astro.42.053102.134034}{\color{magenta}\araa},
  \href{http://adsabs.harvard.edu/abs/2004ARA%26A..42...79B}{42, 79}

\bibitem[{Cain {et~al.}(2021)Cain, D'Aloisio, Gangolli, \&
  Becker}]{cain_short_2021}
Cain, C., D'Aloisio, A., Gangolli, N., \& Becker, G.~D.
  \href{http://adsabs.harvard.edu/abs/2021arXiv210510511C}{2021,
  arXiv:2105.10511}

\bibitem[{Carilli {et~al.}(2007)Carilli, Neri, Wang, Cox, Bertoldi, Walter,
  Fan, Menten, Wagg, Maiolino, Omont, Strauss, Riechers, Lo, Bolatto, \&
  Scoville}]{carilli_detection_2007}
Carilli, C.~L., Neri, R., Wang, R., {et~al.} 2007,
  \href{http://dx.doi.org/10.1086/521648}{\color{magenta}\apj},
  \href{http://adsabs.harvard.edu/abs/2007ApJ...666L...9C}{666, L9}

\bibitem[{Carnall(2017)}]{carnall_spectres_2017}
Carnall, A.~C. \href{http://adsabs.harvard.edu/abs/2017arXiv170505165C}{2017,
  arXiv:1705.05165}

\bibitem[{Chardin {et~al.}(2017)Chardin, Puchwein, \&
  Haehnelt}]{chardin_large-scale_2017}
Chardin, J., Puchwein, E., \& Haehnelt, M.~G. 2017,
  \href{http://dx.doi.org/10.1093/mnras/stw2943}{\color{magenta}\mnras},
  \href{http://adsabs.harvard.edu/abs/2017MNRAS.465.3429C}{465, 3429}

\bibitem[{Chen {et~al.}(2017)Chen, Simcoe, Torrey, Ba{\~n}ados, Cooksey,
  Cooper, Furesz, Matejek, Miller, Turner, Venemans, Decarli, Farina,
  Mazzucchelli, \& Walter}]{chen_mg_2017}
Chen, S.-F.~S., Simcoe, R.~A., Torrey, P., {et~al.} 2017,
  \href{http://dx.doi.org/10.3847/1538-4357/aa9707}{\color{magenta}\apj},
  \href{http://adsabs.harvard.edu/abs/2017ApJ...850..188C}{850, 188}

\bibitem[{Choudhury {et~al.}(2021)Choudhury, Paranjape, \&
  Bosman}]{choudhury_studying_2021}
Choudhury, T.~R., Paranjape, A., \& Bosman, S. E.~I. 2021,
  \href{http://dx.doi.org/10.1093/mnras/stab045}{\color{magenta}\mnras},
  \href{http://adsabs.harvard.edu/abs/2021MNRAS.501.5782C}{501, 5782}

\bibitem[{D'Aloisio {et~al.}(2015)D'Aloisio, McQuinn, \&
  Trac}]{daloisio_large_2015}
D'Aloisio, A., McQuinn, M., \& Trac, H. 2015,
  \href{http://dx.doi.org/10.1088/2041-8205/813/2/L38}{\color{magenta}\apjl},
  \href{http://adsabs.harvard.edu/abs/2015ApJ...813L..38D}{813, L38}

\bibitem[{Davies {et~al.}(2021)Davies, Bosman, Furlanetto, Becker, \&
  D'Aloisio}]{davies_predicament_2021}
Davies, F.~B., Bosman, S. E.~I., Furlanetto, S.~R., Becker, G.~D., \&
  D'Aloisio, A. \href{http://adsabs.harvard.edu/abs/2021arXiv210510518D}{2021,
  arXiv:2105.10518}

\bibitem[{Davies \& Furlanetto(2016)}]{davies_large_2016}
Davies, F.~B. \& Furlanetto, S.~R. 2016,
  \href{http://dx.doi.org/10.1093/mnras/stw931}{\color{magenta}\mnras},
  \href{http://adsabs.harvard.edu/abs/2016MNRAS.460.1328D}{460, 1328}

\bibitem[{Davies {et~al.}(2018{\natexlab{a}})Davies, Hennawi, Ba{\~n}ados,
  Luki{\'c}, Decarli, Fan, Farina, Mazzucchelli, Rix, Venemans, Walter, Wang,
  \& Yang}]{davies_quantitative_2018}
Davies, F.~B., Hennawi, J.~F., Ba{\~n}ados, E., {et~al.} 2018{\natexlab{a}},
  \href{http://dx.doi.org/10.3847/1538-4357/aad6dc}{\color{magenta}\apj},
  \href{http://adsabs.harvard.edu/abs/2018ApJ...864..142D}{864, 142}

\bibitem[{Davies {et~al.}(2018{\natexlab{b}})Davies, Hennawi, Ba{\~n}ados,
  Simcoe, Decarli, Fan, Farina, Mazzucchelli, Rix, Venemans, Walter, Wang, \&
  Yang}]{davies_predicting_2018}
---. 2018{\natexlab{b}},
  \href{http://dx.doi.org/10.3847/1538-4357/aad7f8}{\color{magenta}\apj},
  \href{http://adsabs.harvard.edu/abs/2018ApJ...864..143D}{864, 143}

\bibitem[{Dayal \& Ferrara(2018)}]{dayal_early_2018}
Dayal, P. \& Ferrara, A. 2018,
  \href{http://dx.doi.org/10.1016/j.physrep.2018.10.002}{\color{magenta}\physrep},
  \href{http://adsabs.harvard.edu/abs/2018PhR...780....1D}{780, 1}

\bibitem[{{de Belsunce} {et~al.}(2021){de Belsunce}, Gratton, Coulton, \&
  Efstathiou}]{de_belsunce_inference_2021}
{de Belsunce}, R., Gratton, S., Coulton, W., \& Efstathiou, G.
  \href{http://adsabs.harvard.edu/abs/2021arXiv210314378D}{2021,
  arXiv:2103.14378}

\bibitem[{Decarli {et~al.}(2018)Decarli, Walter, Venemans, Ba{\~n}ados,
  Bertoldi, Carilli, Fan, Farina, Mazzucchelli, Riechers, Rix, Strauss, Wang,
  \& Yang}]{decarli_alma_2018}
Decarli, R., Walter, F., Venemans, B.~P., {et~al.} 2018,
  \href{http://dx.doi.org/10.3847/1538-4357/aaa5aa}{\color{magenta}\apj},
  \href{http://adsabs.harvard.edu/abs/2018ApJ...854...97D}{854, 97}

\bibitem[{Eilers {et~al.}(2018)Eilers, Davies, \&
  Hennawi}]{eilers_opacity_2018}
Eilers, A.-C., Davies, F.~B., \& Hennawi, J.~F. 2018,
  \href{http://dx.doi.org/10.3847/1538-4357/aad4fd}{\color{magenta}\apj},
  \href{http://adsabs.harvard.edu/abs/2018ApJ...864...53E}{864, 53}

\bibitem[{Eilers {et~al.}(2017)Eilers, Davies, Hennawi, Prochaska, Luki{\'c},
  \& Mazzucchelli}]{eilers_implications_2017}
Eilers, A.-C., Davies, F.~B., Hennawi, J.~F., {et~al.} 2017,
  \href{http://dx.doi.org/10.3847/1538-4357/aa6c60}{\color{magenta}\apj},
  \href{http://adsabs.harvard.edu/abs/2017ApJ...840...24E}{840, 24}

\bibitem[{Eilers {et~al.}(2020)Eilers, Hennawi, Decarli, Davies, Venemans,
  Walter, Ba{\~n}ados, Fan, Farina, Mazzucchelli, Novak, Schindler, Simcoe,
  Wang, \& Yang}]{eilers_detecting_2020}
Eilers, A.-C., Hennawi, J.~F., Decarli, R., {et~al.} 2020,
  \href{http://dx.doi.org/10.3847/1538-4357/aba52e}{\color{magenta}\apj},
  \href{http://adsabs.harvard.edu/abs/2020ApJ...900...37E}{900, 37}

\bibitem[{Fan {et~al.}(2006)Fan, Strauss, Becker, White, Gunn, Knapp, Richards,
  Schneider, Brinkmann, \& Fukugita}]{fan_constraining_2006}
Fan, X., Strauss, M.~A., Becker, R.~H., {et~al.} 2006,
  \href{http://dx.doi.org/10.1086/504836}{\color{magenta}\aj},
  \href{http://adsabs.harvard.edu/abs/2006AJ....132..117F/abstract}{132, 117}

\bibitem[{Furlanetto {et~al.}(2004)Furlanetto, Hernquist, \&
  Zaldarriaga}]{furlanetto_constraining_2004}
Furlanetto, S.~R., Hernquist, L., \& Zaldarriaga, M. 2004,
  \href{http://dx.doi.org/10.1111/j.1365-2966.2004.08225.x}{\color{magenta}\mnras},
  \href{http://adsabs.harvard.edu/abs/2004MNRAS.354..695F}{354, 695}

\bibitem[{Gaikwad {et~al.}(2020)Gaikwad, Srianand, Haehnelt, \&
  Choudhury}]{gaikwad_consistent_2020}
Gaikwad, P., Srianand, R., Haehnelt, M.~G., \& Choudhury, T.~R.
  \href{http://adsabs.harvard.edu/abs/2020arXiv200900016G}{2020,
  arXiv:2009.00016}

\bibitem[{Gallerani {et~al.}(2008)Gallerani, Ferrara, Fan, \&
  Choudhury}]{gallerani_glimpsing_2008}
Gallerani, S., Ferrara, A., Fan, X., \& Choudhury, T.~R. 2008,
  \href{http://dx.doi.org/10.1111/j.1365-2966.2008.13029.x}{\color{magenta}\mnras},
  \href{http://adsabs.harvard.edu/abs/2008MNRAS.386..359G}{386, 359}

\bibitem[{Greig {et~al.}(2019)Greig, Mesinger, \&
  Ba{\~n}ados}]{greig_constraints_2019}
Greig, B., Mesinger, A., \& Ba{\~n}ados, E. 2019,
  \href{http://dx.doi.org/10.1093/mnras/stz230}{\color{magenta}\mnras},
  \href{http://adsabs.harvard.edu/abs/2019MNRAS.484.5094G}{484, 5094}

\bibitem[{Greig {et~al.}(2017)Greig, Mesinger, Haiman, \&
  Simcoe}]{greig_are_2017}
Greig, B., Mesinger, A., Haiman, Z., \& Simcoe, R.~A. 2017,
  \href{http://dx.doi.org/10.1093/mnras/stw3351}{\color{magenta}\mnras},
  \href{http://adsabs.harvard.edu/abs/2017MNRAS.466.4239G}{466, 4239}

\bibitem[{Haardt \& Madau(2012)}]{haardt_radiative_2012}
Haardt, F. \& Madau, P. 2012,
  \href{http://dx.doi.org/10.1088/0004-637X/746/2/125}{\color{magenta}\apj},
  \href{http://adsabs.harvard.edu/abs/2012ApJ...746..125H}{746, 125}

\bibitem[{Harris {et~al.}(2020)Harris, Millman, {van der Walt}, Gommers,
  Virtanen, Cournapeau, Wieser, Taylor, Berg, Smith, Kern, Picus, Hoyer, {van
  Kerkwijk}, Brett, Haldane, {del R{\'i}o}, Wiebe, Peterson,
  {G{\'e}rard-Marchant}, Sheppard, Reddy, Weckesser, Abbasi, Gohlke, \&
  Oliphant}]{harris_array_2020}
Harris, C.~R., Millman, K.~J., {van der Walt}, S.~J., {et~al.} 2020,
  \href{http://dx.doi.org/10.1038/s41586-020-2649-2}{\color{magenta}\nat},
  \href{http://adsabs.harvard.edu/abs/2020Natur.585..357H}{585, 357}

\bibitem[{Hoag {et~al.}(2019)Hoag, Brada{\v c}, Huang, Mason, Treu, Schmidt,
  Trenti, Strait, Lemaux, Finney, \& Paddock}]{hoag_constraining_2019}
Hoag, A., Brada{\v c}, M., Huang, K., {et~al.} 2019,
  \href{http://dx.doi.org/10.3847/1538-4357/ab1de7}{\color{magenta}\apj},
  \href{http://adsabs.harvard.edu/abs/2019ApJ...878...12H}{878, 12}

\bibitem[{Horne(1986)}]{horne_optimal_1986}
Horne, K. 1986, \href{http://dx.doi.org/10.1086/131801}{\color{magenta}\pasp},
  \href{http://adsabs.harvard.edu/abs/1986PASP...98..609H}{98, 609}

\bibitem[{Hu {et~al.}(2019)Hu, Wang, Zheng, Malhotra, Rhoads, Infante,
  Barrientos, Yang, Jiang, Kang, Perez, Wold, Hibon, Jiang, Khostovan, Valdes,
  Walker, Galaz, Coughlin, Harish, Kong, Pharo, \& Zheng}]{hu_ly_2019}
Hu, W., Wang, J., Zheng, Z.-Y., {et~al.} 2019,
  \href{http://dx.doi.org/10.3847/1538-4357/ab4cf4}{\color{magenta}\apj},
  \href{http://adsabs.harvard.edu/abs/2019ApJ...886...90H}{886, 90}

\bibitem[{Hunter(2007)}]{hunter_matplotlib_2007}
Hunter, J.~D. 2007,
  \href{http://dx.doi.org/10.1109/MCSE.2007.55}{\color{magenta}CSE},
  \href{http://adsabs.harvard.edu/abs/2007CSE.....9...90H}{9, 90}

\bibitem[{Jiang {et~al.}(2007)Jiang, Fan, Vestergaard, Kurk, Walter, Kelly, \&
  Strauss}]{jiang_gemini_2007}
Jiang, L., Fan, X., Vestergaard, M., {et~al.} 2007,
  \href{http://dx.doi.org/10.1086/520811}{\color{magenta}\aj},
  \href{http://adsabs.harvard.edu/abs/2007AJ....134.1150J}{134, 1150}

\bibitem[{Jones {et~al.}(2013)Jones, Noll, Kausch, Szyszka, \&
  Kimeswenger}]{jones_advanced_2013}
Jones, A., Noll, S., Kausch, W., Szyszka, C., \& Kimeswenger, S. 2013,
  \href{http://dx.doi.org/10.1051/0004-6361/201322433}{\color{magenta}\aap},
  \href{http://adsabs.harvard.edu/abs/2013A%26A...560A..91J}{560, A91}

\bibitem[{Kashino {et~al.}(2020)Kashino, Lilly, Shibuya, Ouchi, \&
  Kashikawa}]{kashino_evidence_2020}
Kashino, D., Lilly, S.~J., Shibuya, T., Ouchi, M., \& Kashikawa, N. 2020,
  \href{http://dx.doi.org/10.3847/1538-4357/ab5a7d}{\color{magenta}\apj},
  \href{http://adsabs.harvard.edu/abs/2020ApJ...888....6K}{888, 6}

\bibitem[{Keating {et~al.}(2020{\natexlab{a}})Keating, Kulkarni, Haehnelt,
  Chardin, \& Aubert}]{keating_constraining_2020}
Keating, L.~C., Kulkarni, G., Haehnelt, M.~G., Chardin, J., \& Aubert, D.
  2020{\natexlab{a}},
  \href{http://dx.doi.org/10.1093/mnras/staa1909}{\color{magenta}\mnras},
  \href{http://adsabs.harvard.edu/abs/2020MNRAS.497..906K}{497, 906}

\bibitem[{Keating {et~al.}(2020{\natexlab{b}})Keating, Weinberger, Kulkarni,
  Haehnelt, Chardin, \& Aubert}]{keating_long_2020}
Keating, L.~C., Weinberger, L.~H., Kulkarni, G., {et~al.} 2020{\natexlab{b}},
  \href{http://dx.doi.org/10.1093/mnras/stz3083}{\color{magenta}\mnras},
  \href{http://adsabs.harvard.edu/abs/2020MNRAS.491.1736K}{491, 1736}

\bibitem[{Kelson(2003)}]{kelson_optimal_2003}
Kelson, D.~D. 2003,
  \href{http://dx.doi.org/10.1086/375502}{\color{magenta}\pasp},
  \href{http://adsabs.harvard.edu/abs/2003PASP..115..688K}{115, 688}

\bibitem[{Kulkarni {et~al.}(2019{\natexlab{a}})Kulkarni, Keating, Haehnelt,
  Bosman, Puchwein, Chardin, \& Aubert}]{kulkarni_large_2019}
Kulkarni, G., Keating, L.~C., Haehnelt, M.~G., {et~al.} 2019{\natexlab{a}},
  \href{http://dx.doi.org/10.1093/mnrasl/slz025}{\color{magenta}\mnras},
  \href{http://adsabs.harvard.edu/abs/2019MNRAS.485L..24K/abstract}{485, L24}

\bibitem[{Kulkarni {et~al.}(2019{\natexlab{b}})Kulkarni, Worseck, \&
  Hennawi}]{kulkarni_evolution_2019}
Kulkarni, G., Worseck, G., \& Hennawi, J.~F. 2019{\natexlab{b}},
  \href{http://dx.doi.org/10.1093/mnras/stz1493}{\color{magenta}\mnras},
  \href{http://adsabs.harvard.edu/abs/2019MNRAS.488.1035K}{488, 1035}

\bibitem[{Kurk {et~al.}(2007)Kurk, Walter, Fan, Jiang, Riechers, Rix,
  Pentericci, Strauss, Carilli, \& Wagner}]{kurk_black_2007}
Kurk, J.~D., Walter, F., Fan, X., {et~al.} 2007,
  \href{http://dx.doi.org/10.1086/521596}{\color{magenta}\apj},
  \href{http://adsabs.harvard.edu/abs/2007ApJ...669...32K}{669, 32}

\bibitem[{Maiolino {et~al.}(2005)Maiolino, Cox, Caselli, Beelen, Bertoldi,
  Carilli, Kaufman, Menten, Nagao, Omont, Wei{\ss}, Walmsley, \&
  Walter}]{maiolino_first_2005}
Maiolino, R., Cox, P., Caselli, P., {et~al.} 2005,
  \href{http://dx.doi.org/10.1051/0004-6361:200500165}{\color{magenta}\aap},
  \href{http://adsabs.harvard.edu/abs/2005A&A...440L..51M/abstract}{440, L51}

\bibitem[{Mason {et~al.}(2019)Mason, Fontana, Treu, Schmidt, Hoag, Abramson,
  Amorin, Brada{\v c}, Guaita, Jones, Henry, Malkan, Pentericci, Trenti, \&
  Vanzella}]{mason_inferences_2019}
Mason, C.~A., Fontana, A., Treu, T., {et~al.} 2019,
  \href{http://dx.doi.org/10.1093/mnras/stz632}{\color{magenta}\mnras},
  \href{http://adsabs.harvard.edu/abs/2019MNRAS.485.3947M}{485, 3947}

\bibitem[{Mason {et~al.}(2018)Mason, Treu, Dijkstra, Mesinger, Trenti,
  Pentericci, {de Barros}, \& Vanzella}]{mason_universe_2018}
Mason, C.~A., Treu, T., Dijkstra, M., {et~al.} 2018,
  \href{http://dx.doi.org/10.3847/1538-4357/aab0a7}{\color{magenta}\apj},
  \href{http://adsabs.harvard.edu/abs/2018ApJ...856....2M}{856, 2}

\bibitem[{Mazzucchelli {et~al.}(2017)Mazzucchelli, Ba{\~n}ados, Venemans,
  Decarli, Farina, Walter, Eilers, Rix, Simcoe, Stern, Fan, Schlafly, De~Rosa,
  Hennawi, Chambers, Greiner, Burgett, Draper, Kaiser, Kudritzki, Magnier,
  Metcalfe, Waters, \& Wainscoat}]{mazzucchelli_physical_2017}
Mazzucchelli, C., Ba{\~n}ados, E., Venemans, B.~P., {et~al.} 2017,
  \href{http://dx.doi.org/10.3847/1538-4357/aa9185}{\color{magenta}\apj},
  \href{http://adsabs.harvard.edu/abs/2017ApJ...849...91M}{849, 91}

\bibitem[{McGreer {et~al.}(2015)McGreer, Mesinger, \&
  D'Odorico}]{mcgreer_model-independent_2015}
McGreer, I.~D., Mesinger, A., \& D'Odorico, V. 2015,
  \href{http://dx.doi.org/10.1093/mnras/stu2449}{\color{magenta}\mnras},
  \href{http://adsabs.harvard.edu/abs/2015MNRAS.447..499M}{447, 499}

\bibitem[{McQuinn(2016)}]{mcquinn_evolution_2016}
McQuinn, M. 2016,
  \href{http://dx.doi.org/10.1146/annurev-astro-082214-122355}{\color{magenta}\araa},
  \href{http://adsabs.harvard.edu/abs/2016ARA%26A..54..313M/abstract}{54, 313}

\bibitem[{Meiksin(2020)}]{meiksin_influence_2020}
Meiksin, A. 2020,
  \href{http://dx.doi.org/10.1093/mnras/stz3395}{\color{magenta}\mnras},
  \href{http://adsabs.harvard.edu/abs/2020MNRAS.491.4884M}{491, 4884}

\bibitem[{Nasir \& D'Aloisio(2020)}]{nasir_observing_2020}
Nasir, F. \& D'Aloisio, A. 2020,
  \href{http://dx.doi.org/10.1093/mnras/staa894}{\color{magenta}\mnras},
  \href{http://adsabs.harvard.edu/abs/2020MNRAS.494.3080N}{494, 3080}

\bibitem[{Noll {et~al.}(2012)Noll, Kausch, Barden, Jones, Szyszka, Kimeswenger,
  \& Vinther}]{noll_atmospheric_2012}
Noll, S., Kausch, W., Barden, M., {et~al.} 2012,
  \href{http://dx.doi.org/10.1051/0004-6361/201219040}{\color{magenta}\aap},
  \href{http://adsabs.harvard.edu/abs/2012A%26A...543A..92N}{543, A92}

\bibitem[{Parsa {et~al.}(2018)Parsa, Dunlop, \& McLure}]{parsa_no_2018}
Parsa, S., Dunlop, J.~S., \& McLure, R.~J. 2018,
  \href{http://dx.doi.org/10.1093/mnras/stx2887}{\color{magenta}\mnras},
  \href{http://adsabs.harvard.edu/abs/2018MNRAS.474.2904P}{474, 2904}

\bibitem[{Paschos \& Norman(2005)}]{paschos_statistical_2005}
Paschos, P. \& Norman, M.~L. 2005,
  \href{http://dx.doi.org/10.1086/431787}{\color{magenta}\apj},
  \href{http://adsabs.harvard.edu/abs/2005ApJ...631...59P}{631, 59}

\bibitem[{{Planck Collaboration} {et~al.}(2020){Planck Collaboration}, Aghanim,
  Akrami, Ashdown, Aumont, Baccigalupi, Ballardini, Banday, Barreiro, Bartolo,
  Basak, Battye, Benabed, Bernard, Bersanelli, Bielewicz, Bock, Bond, Borrill,
  Bouchet, Boulanger, Bucher, Burigana, Butler, Calabrese, Cardoso, Carron,
  Challinor, Chiang, Chluba, Colombo, Combet, Contreras, Crill, Cuttaia, {de
  Bernardis}, {de Zotti}, Delabrouille, Delouis, Di~Valentino, Diego, Dor{\'e},
  Douspis, Ducout, Dupac, Dusini, Efstathiou, Elsner, En{\ss}lin, Eriksen,
  Fantaye, Farhang, Fergusson, {Fernandez-Cobos}, Finelli, Forastieri, Frailis,
  Fraisse, Franceschi, Frolov, Galeotta, Galli, Ganga, {G{\'e}nova-Santos},
  Gerbino, Ghosh, {Gonz{\'a}lez-Nuevo}, G{\'o}rski, Gratton, Gruppuso,
  Gudmundsson, Hamann, Handley, Hansen, Herranz, Hildebrandt, Hivon, Huang,
  Jaffe, Jones, Karakci, Keih{\"a}nen, Keskitalo, Kiiveri, Kim, Kisner, Knox,
  Krachmalnicoff, Kunz, {Kurki-Suonio}, Lagache, Lamarre, Lasenby, Lattanzi,
  Lawrence, Le~Jeune, Lemos, Lesgourgues, Levrier, Lewis, Liguori, Lilje,
  Lilley, Lindholm, {L{\'o}pez-Caniego}, Lubin, Ma, {Mac{\'i}as-P{\'e}rez},
  Maggio, Maino, Mandolesi, Mangilli, {Marcos-Caballero}, Maris, Martin,
  Martinelli, {Mart{\'i}nez-Gonz{\'a}lez}, Matarrese, Mauri, McEwen, Meinhold,
  Melchiorri, Mennella, Migliaccio, Millea, Mitra, {Miville-Desch{\^e}nes},
  Molinari, Montier, Morgante, Moss, Natoli, {N{\o}rgaard-Nielsen}, Pagano,
  Paoletti, Partridge, Patanchon, Peiris, Perrotta, Pettorino, Piacentini,
  Polastri, Polenta, Puget, Rachen, Reinecke, Remazeilles, Renzi, Rocha,
  Rosset, Roudier, {Rubi{\~n}o-Mart{\'i}n}, {Ruiz-Granados}, Salvati, Sandri,
  Savelainen, Scott, Shellard, Sirignano, Sirri, Spencer, Sunyaev, {Suur-Uski},
  Tauber, Tavagnacco, Tenti, Toffolatti, Tomasi, Trombetti, Valenziano,
  Valiviita, Van~Tent, Vibert, Vielva, Villa, Vittorio, Wandelt, Wehus, White,
  White, Zacchei, \& Zonca}]{planck_collaboration_planck_2020}
{Planck Collaboration}, Aghanim, N., Akrami, Y., {et~al.} 2020,
  \href{http://dx.doi.org/10.1051/0004-6361/201833910}{\color{magenta}\aap},
  \href{http://adsabs.harvard.edu/abs/2020A%26A...641A...6P}{641, A6}

\bibitem[{Qin {et~al.}(2021)Qin, Mesinger, Bosman, \&
  Viel}]{qin_reionization_2021}
Qin, Y., Mesinger, A., Bosman, S. E.~I., \& Viel, M. 2021,
  \href{http://adsabs.harvard.edu/abs/2021arXiv210109033Q}{2101,
  arXiv:2101.09033}

\bibitem[{Sheinis {et~al.}(2002)Sheinis, Bolte, Epps, Kibrick, Miller, Radovan,
  Bigelow, \& Sutin}]{sheinis_esi_2002}
Sheinis, A.~I., Bolte, M., Epps, H.~W., {et~al.} 2002,
  \href{http://dx.doi.org/10.1086/341706}{\color{magenta}\pasp},
  \href{http://adsabs.harvard.edu/abs/2002PASP..114..851S}{114, 851}

\bibitem[{Shen {et~al.}(2019)Shen, Wu, Jiang, Ba{\~n}ados, Fan, Ho, Riechers,
  Strauss, Venemans, Vestergaard, Walter, Wang, Willott, Wu, \&
  Yang}]{shen_gemini_2019}
Shen, Y., Wu, J., Jiang, L., {et~al.} 2019,
  \href{http://dx.doi.org/10.3847/1538-4357/ab03d9}{\color{magenta}\apj},
  \href{http://adsabs.harvard.edu/abs/2019ApJ...873...35S}{873, 35}

\bibitem[{Songaila \& Cowie(2002)}]{songaila_approaching_2002}
Songaila, A. \& Cowie, L.~L. 2002,
  \href{http://dx.doi.org/10.1086/340079}{\color{magenta}\aj},
  \href{http://adsabs.harvard.edu/abs/2002AJ....123.2183S}{123, 2183}

\bibitem[{Springel(2005)}]{springel_cosmological_2005}
Springel, V. 2005,
  \href{http://dx.doi.org/10.1111/j.1365-2966.2005.09655.x}{\color{magenta}\mnras},
  \href{http://adsabs.harvard.edu/abs/2005MNRAS.364.1105S}{364, 1105}

\bibitem[{Trac \& Pen(2004)}]{trac_moving_2004}
Trac, H. \& Pen, U.-L. 2004,
  \href{http://dx.doi.org/10.1016/j.newast.2004.02.002}{\color{magenta}\na},
  \href{http://adsabs.harvard.edu/abs/2004NewA....9..443T}{9, 443}

\bibitem[{Venemans {et~al.}(2020)Venemans, Walter, Neeleman, Novak, Otter,
  Decarli, Ba{\~n}ados, Drake, Farina, Kaasinen, Mazzucchelli, Carilli, Fan,
  Rix, \& Wang}]{venemans_kiloparsec-scale_2020-1}
Venemans, B.~P., Walter, F., Neeleman, M., {et~al.} 2020,
  \href{http://dx.doi.org/10.3847/1538-4357/abc563}{\color{magenta}\apj},
  \href{http://adsabs.harvard.edu/abs/2020ApJ...904..130V}{904, 130}

\bibitem[{Vernet {et~al.}(2011)Vernet, Dekker, D'Odorico, Kaper, Kjaergaard,
  Hammer, Randich, Zerbi, Groot, Hjorth, Guinouard, Navarro, Adolfse, Albers,
  Amans, Andersen, Andersen, Binetruy, Bristow, Castillo, Chemla, Christensen,
  Conconi, Conzelmann, Dam, {de Caprio}, {de Ugarte Postigo}, Delabre, {di
  Marcantonio}, Downing, Elswijk, Finger, Fischer, Flores, Fran{\c c}ois,
  Goldoni, Guglielmi, Haigron, Hanenburg, Hendriks, Horrobin, Horville, Jessen,
  Kerber, Kern, Kiekebusch, Kleszcz, Klougart, Kragt, Larsen, Lizon, Lucuix,
  Mainieri, Manuputy, Martayan, Mason, Mazzoleni, Michaelsen, Modigliani,
  Moehler, M{\o}ller, Norup~S{\o}rensen, N{\o}rregaard, P{\'e}roux, Patat,
  Pena, Pragt, Reinero, Rigal, Riva, Roelfsema, Royer, Sacco, Santin,
  Schoenmaker, Spano, Sweers, Ter~Horst, Tintori, Tromp, {van Dael}, {van der
  Vliet}, Venema, Vidali, Vinther, Vola, Winters, Wistisen, Wulterkens, \&
  Zacchei}]{vernet_x-shooter_2011}
Vernet, J., Dekker, H., D'Odorico, S., {et~al.} 2011,
  \href{http://dx.doi.org/10.1051/0004-6361/201117752}{\color{magenta}\aap},
  \href{http://adsabs.harvard.edu/abs/2011A%26A...536A.105V}{536, A105}

\bibitem[{Walther {et~al.}(2019)Walther, O{\~n}orbe, Hennawi, \&
  Luki{\'c}}]{walther_new_2019}
Walther, M., O{\~n}orbe, J., Hennawi, J.~F., \& Luki{\'c}, Z. 2019,
  \href{http://dx.doi.org/10.3847/1538-4357/aafad1}{\color{magenta}\apj},
  \href{http://adsabs.harvard.edu/abs/2019ApJ...872...13W}{872, 13}

\bibitem[{Wang {et~al.}(2020)Wang, Davies, Yang, Hennawi, Fan, Barth, Jiang,
  Wu, Mudd, Ba{\~n}ados, Bian, Decarli, Eilers, Farina, Venemans, Walter, \&
  Yue}]{wang_significantly_2020-1}
Wang, F., Davies, F.~B., Yang, J., {et~al.} 2020,
  \href{http://dx.doi.org/10.3847/1538-4357/ab8c45}{\color{magenta}\apj},
  \href{http://adsabs.harvard.edu/abs/2020ApJ...896...23W}{896, 23}

\bibitem[{Wang {et~al.}(2021)Wang, Fan, Yang, Mazzucchelli, Wu, Li,
  Ba{\~n}ados, Farina, Nanni, Ai, Bian, Davies, Decarli, Hennawi, Schindler,
  Venemans, \& Walter}]{wang_revealing_2021}
Wang, F., Fan, X., Yang, J., {et~al.} 2021,
  \href{http://dx.doi.org/10.3847/1538-4357/abcc5e}{\color{magenta}\apj},
  \href{http://adsabs.harvard.edu/abs/2021ApJ...908...53W}{908, 53}

\bibitem[{Wang {et~al.}(2019)Wang, Wang, Fan, Wu, Yang, Neri, \&
  Yue}]{wang_spatially_2019}
Wang, F., Wang, R., Fan, X., {et~al.} 2019,
  \href{http://dx.doi.org/10.3847/1538-4357/ab2717}{\color{magenta}\apj},
  \href{http://adsabs.harvard.edu/abs/2019ApJ...880....2W}{880, 2}

\bibitem[{Wang {et~al.}(2010)Wang, Carilli, Neri, Riechers, Wagg, Walter,
  Bertoldi, Menten, Omont, Cox, \& Fan}]{wang_molecular_2010}
Wang, R., Carilli, C.~L., Neri, R., {et~al.} 2010,
  \href{http://dx.doi.org/10.1088/0004-637X/714/1/699}{\color{magenta}\apj},
  \href{http://adsabs.harvard.edu/abs/2010ApJ...714..699W}{714, 699}

\bibitem[{Wang {et~al.}(2013)Wang, Wagg, Carilli, Walter, Lentati, Fan,
  Riechers, Bertoldi, Narayanan, Strauss, Cox, Omont, Menten, Knudsen, Neri, \&
  Jiang}]{wang_star_2013}
Wang, R., Wagg, J., Carilli, C.~L., {et~al.} 2013,
  \href{http://dx.doi.org/10.1088/0004-637X/773/1/44}{\color{magenta}\apj},
  \href{http://adsabs.harvard.edu/abs/2013ApJ...773...44W}{773, 44}

\bibitem[{Wold {et~al.}(2021)Wold, Malhotra, Rhoads, Wang, Hu, Perez, Zheng,
  Khostovan, Walker, Barrientos, {Gonz{\'a}lez-L{\'o}pez}, Harish, Infante,
  Jiang, Pharo, {Moya-Sierralta}, Valdes, \& Yang}]{wold_lager_2021-1}
Wold, I. G.~B., Malhotra, S., Rhoads, J., {et~al.}
  \href{http://adsabs.harvard.edu/abs/2021arXiv210512191W}{2021,
  arXiv:2105.12191}

\bibitem[{Yang {et~al.}(2019)Yang, Venemans, Wang, Fan, Novak, Decarli, Walter,
  Yue, Momjian, Keeton, Wang, Zabludoff, Wu, \& Bian}]{yang_far-infrared_2019}
Yang, J., Venemans, B., Wang, F., {et~al.} 2019,
  \href{http://dx.doi.org/10.3847/1538-4357/ab2a02}{\color{magenta}\apj},
  \href{http://adsabs.harvard.edu/abs/2019ApJ...880..153Y}{880, 153}

\bibitem[{Yang {et~al.}(2020{\natexlab{a}})Yang, Wang, Fan, Hennawi, Davies,
  Yue, Banados, Wu, Venemans, Barth, Bian, Boutsia, Decarli, Farina, Green,
  Jiang, Li, Mazzucchelli, \& Walter}]{yang_poniuaena_2020}
Yang, J., Wang, F., Fan, X., {et~al.} 2020{\natexlab{a}},
  \href{http://dx.doi.org/10.3847/2041-8213/ab9c26}{\color{magenta}\apjl},
  \href{http://adsabs.harvard.edu/abs/2020ApJ...897L..14Y}{897, L14}

\bibitem[{Yang {et~al.}(2020{\natexlab{b}})Yang, Wang, Fan, Hennawi, Davies,
  Yue, Eilers, Farina, Wu, Bian, Pacucci, \& Lee}]{yang_measurements_2020-1}
---. 2020{\natexlab{b}},
  \href{http://dx.doi.org/10.3847/1538-4357/abbc1b}{\color{magenta}\apj},
  \href{http://adsabs.harvard.edu/abs/2020ApJ...904...26Y}{904, 26}

\end{thebibliography}
\end{document}